\pdfoutput=1
\documentclass[11pt,aps,a4paper,eqsecnum,amsmath,amssymb,longbibliography,notitlepage,nofootinbib]{revtex4-1}

\usepackage[mathscr]{eucal}
\usepackage{amsthm,amssymb}
\usepackage{hhline}
\usepackage{psfrag}
\usepackage{mathrsfs} 
\usepackage{hyperref}
\usepackage{array}
\usepackage{multirow}
\usepackage{textcomp}
\usepackage{hyperref}
\hypersetup{
    colorlinks,
    linkcolor={cyan!40!black},
    citecolor={green!40!black},
    urlcolor={blue!50!black}
}
\usepackage{mathrsfs} 
\usepackage{enumitem}
\usepackage{xcolor}

\DeclareMathAlphabet{\mathpzc}{OT1}{pzc}{m}{it}

\definecolor{byzantine}{rgb}{0.74, 0.2, 0.64}

\DeclareFontFamily{U}{BOONDOX-calo}{\skewchar\font=45 }
\DeclareFontShape{U}{BOONDOX-calo}{m}{n}{
  <-> s*[1.05] BOONDOX-r-calo}{}
\DeclareFontShape{U}{BOONDOX-calo}{b}{n}{
  <-> s*[1.05] BOONDOX-b-calo}{}
\DeclareMathAlphabet{\mathcalboondox}{U}{BOONDOX-calo}{m}{n}
\SetMathAlphabet{\mathcalboondox}{bold}{U}{BOONDOX-calo}{b}{n}
\DeclareMathAlphabet{\mathbcalboondox}{U}{BOONDOX-calo}{b}{n}

\usepackage{graphicx}
\graphicspath{ {./images/} }
\usepackage{bbm}
\usepackage{bm}

\usepackage{tikz}
\usetikzlibrary{decorations.markings}
\usepackage{mathtools}
\usepackage{float}
\usepackage{subfigure}
\usepackage{braket}
\makeatletter
\def\@bibdataout@aps{%
\immediate\write\@bibdataout{%
@CONTROL{%
apsrev41Control%
\longbibliography@sw{%
    ,author="08",editor="1",pages="1",title="0",year="1"%
    }{%
    ,author="08",editor="1",pages="1",title="",year="1"%
    }%
  }%
}%
\if@filesw \immediate \write \@auxout {\string \citation {apsrev41Control}}\fi 
}
\makeatother

\newcommand{\kommu}[2]{\left[#1,#2 \right] }

\newcommand{\tr}[0]{\text{Tr}}
\newcommand{\ri}[0]{{\rm i}}
\newcommand{\ecelll}[3]{
\draw[<-,thick] (#1,#2+0.5) -- (#1,#2-0.5);
\draw[<-,thick] (#1-1,#2) -- (#1+1,#2);
\node[anchor=south west] at (#1+0.1,#2+0.1) {\footnotesize#3};
}

\newcommand{\cross}[2]{
\draw[-,thick](#1-0.1,#2-0.1)--(#1+0.1,#2+0.1);
\draw[-,thick] (#1-0.1,#2+0.1)--(#1+0.1,#2-0.1);
}
\newcommand{\deri}[1]{\frac{\text{d}}{\text{d}#1}}
\newcommand{\re}{\mbox{e}}

\begin{document}

\title{Finite-size spectrum of the staggered six-vertex model with antidiagonal boundary conditions}

\author{Holger Frahm}
\author{Sascha Gehrmann}

\affiliation{Institut f$\ddot{{\rm u}}$r Theoretische Physik, 
Leibniz Universit$\ddot{{\rm a}}$t Hannover\\
Appelstra\ss e 2, 30167 Hannover, }

\date{\today}

\begin{abstract}
The finite-size spectrum of the critical staggered six-vertex model with antidiagonal boundary conditions is studied. Similar to the case of periodic boundary conditions, we identify three different phases. In two of those, the underlying conformal field theory can be identified to be related to the twisted $U(1)$ Kac-Moody algebra. In contrast, the finite size scaling in the third regime, whose critical behaviour with the (quasi-)periodic BCs is related to the 2d black hole CFTs possessing a non-compact degree of freedom, is more subtle. Here with antidiagonal BCs imposed, the corrections to the scaling of the ground state grow logarithmically with the system size, while the energy gaps appear to close logarithmically.  Moreover, we obtain an explicit formula for the Q-operator which is useful for numerical implementation. 
\end{abstract}

\maketitle

\section{Introduction and main results}
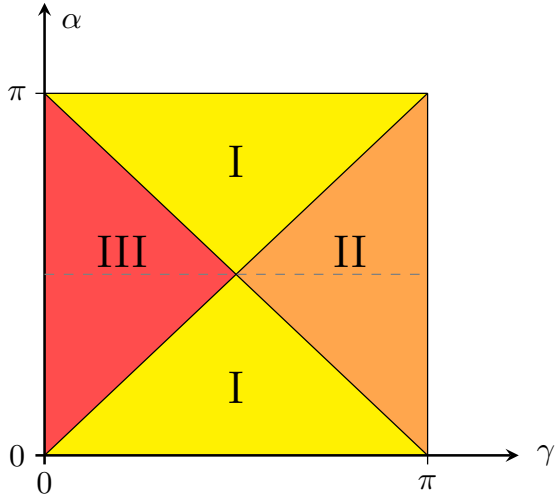
\begin{figure}
    \centering
    \scalebox{1.2}{
\begin{tikzpicture}

\path[fill =green, opacity = 1] (0,0) -- (4.2,0) -- (2.1,2) -- cycle;
\path[fill =yellow, opacity = 1] (0,0) -- (4.2,0) -- (2.1,2) -- cycle;
\path[fill = green, opacity = 1] (0,4) -- (4.2,4) -- (2.1,2) -- cycle;
\path[fill = yellow, opacity = 1] (0,4) -- (4.2,4) -- (2.1,2) -- cycle;
\path[fill = orange, opacity = 0.7] (4.2,0) -- (2.1,2) -- (4.2,4) -- cycle;
\path[fill = red, opacity = 0.7] (0,0) -- (2.1,2) -- (0,4) -- cycle;
\draw[-stealth,thick] (-0.1,0) -> (5.2,0); 
\draw[-stealth,thick] (0,-0.1) -> (0,5); 
\draw (4.2,-0.1) -- (4.2,4);
\draw (-0.1,4) -- (4.2,4);
\draw (0,0) -- (4.2,4);
\draw (0,4) -- (4.2,0);
\draw[dashed,gray] (0,2) -- (4.2,2);
\node at (-0.3,0) {$0$};
\node at (-0.3,4) {$\pi$};
\node at (0,-0.3) {$0$};
\node at (4.2,-0.3) {$\pi$};
\node at (3.35,2.25) {\Large II};
\node at (0.85,2.25) {\Large III};
\node at (2.1,0.75) {\Large I};
\node at (2.1,3.25) {\Large I};
\node at (5.5,0) {$\gamma$};
\node at (0.3,4.8) {$\alpha$};
\end{tikzpicture}}
    \caption{Phase diagram of the staggered six-vertex model with (quasi-)periodic boundary conditions in the critical regime with anisotropy $0<\gamma\leq\pi$ \cite{FrMa12,KoLu23}, lines with fixed staggering parameters $\alpha$ and $\pi-\alpha$ are identified by the duality transformation (\ref{ntjfn}). In phase I the scaling limit is described by a compact free boson -- as in the homogeneous limit $\alpha\to0$. The critical degrees of freedom in phase II are one massless compact boson and two Majorana fermions. In phase III the low energy excitations of the model have been identified with those of the black hole coset model with a non-compact degree of freedom.}
    \label{fig:phases}
\end{figure}

The $\mathbb{Z}_2$-staggered six-vertex model in its critical regime, parameterized by the anisotropy $0<\gamma\leq\pi$, has attracted a lot of attention in recent years.  Much of the prior work has focused on the model with (quasi-)periodic boundary conditions \cite{JaSa06,IkJS08,IkJS10,IkJS12,FrMa12,CaIk13,FrSe14,BKKL19,BKKL21a} where the staggered model has been shown to exhibit several critical phases, depending on the anisotropy and the choice of staggering, see Fig.~\ref{fig:phases}.
In phase~I, i.e.\ for staggering parameter $\alpha$ less than $\min(\gamma,\pi-\gamma)$ (or $\alpha>\max(\gamma,\pi-\gamma)$ related by duality, see Eq.~(\ref{ntjfn}) below), the critical behaviour is that of the homogeneous model, i.e.\ described by a free boson with compactification radius depending on $\gamma$.  In the phases around the 'self-dual' line $\alpha=\pi/2$ the field content of the low energy effective theory depends on the anisotropy: for $\gamma>\pi/2$ (phase~II) the scaling limit of the model is described by a conformal field theory (CFT) consisting of a free compact boson and two Majorana fermions \cite{IkJS10,KoLu2021}. Finally, in phase~III realized for $\gamma<\pi/2$ the critical behaviour is related to the 2d black hole CFTs which possess a continuous component in the conformal spectrum \cite{IkJS12,CaIk13,FrSe14,BKKL21a}.

The appearance of such CFTs in the scaling limit of a lattice model with finitely many degrees of freedom per site has motivated further studies of this model with different boundary conditions (BCs).  In a series of works integrable open boundary conditions leaving the lattice model invariant under the quantum group $U_q(\mathfrak{sl}(2))$ have been studied. Interestingly, two such boundary conditions can be constructed for the lattice model, both influencing the physical properties of the model in phase~III: one of these BCs leads to a purely discrete set of conformal weights leading to a \emph{compact} boundary CFT describing the continuum limit of the model \cite{RoJS19}.  With the second choice for the BCs the symmetry of the lattice model's ground state is spontaneously broken.  The finite size spectrum, however, contains both discrete and continuous parts allowing for the decomposition into irreps of the $\mathcal{W}_\infty$ algebra, the extended conformal symmetry of the model \cite{RoJS21,FrGe22,FrGK23}. While there remain some open questions regarding the relation between this scaling limit and possible D-brane constructions for the black hole CFT both numerical and analytical studies indicate that the two BCs realize an RG flow from an (unstable) non-compact boundary CFT to a (stable) compact one \cite{RoJS21,FrGe23}.
 
Here we continue these studies of the influence of BCs on the critical properties of the staggered six-vertex model by considering anti-diagonal BCs which break the $U(1)$ symmetry of the staggered six-vertex model into discrete ones.  The homogeneous model with such BCs has been solved using Bethe ansatz methods \cite{YuBa95a,BBOY95} and the conformal spectrum in the different symmetry sectors has been found to correspond to a twisted $U(1)$ Kac-Moody algebra and does not depend on the anisotropy $\gamma$ \cite{GeRS88,ABGR88,NiWF09}.  
Our paper is organized as follows: in the following section, we define the model and construct the commuting operators characterizing the spectrum and its symmetries.  In Section~\ref{sec:fs} the finite-size spectrum of the model in the different phases is studied.  To identify the root configurations parameterizing the low energy Bethe states we have generalized the construction of the Baxter Q-operator developed for the homogeneous model \cite{BBOY95} to the inhomogeneous case and present an explicit formula for its matrix elements in an appendix.  Since the latter does not involve any matrix operations it is particularly suitable for an implementation on a computer.  Based on these root configurations we construct RG trajectories to extract the scaling dimensions from the finite size data for the low-lying eigenenergies.  We find that the low energy modes in phases~I and II can be described in terms of twisted $U(1)$ Kac-Moody algebras, similar as in the homogeneous model \cite{GeRS88,ABGR88}.  In phase~III, however, the corrections to scaling along the RG trajectories grow logarithmically with the system size.  Such a behaviour has been observed before for particular states in the periodic model where they have been argued to leave the low energy spectrum and therefore not to be relevant for the scaling limit.  Here \emph{all} low energy states that we have identified for small system sizes show this behaviour.  For the gap between the ground state and the first excitation, however, the corrections to scaling appear to close in the scaling limit.  Larger system sizes need to be considered though to make a quantitative statement on the finite-size scaling.  Therefore the characterization of the scaling limit in this phase is left to a future research project.

\section{Definition of the model}
In this work, we consider the R-matrix of the six-vertex model acting as an endomorphism on $V_i \otimes V_j$ with $V\sim \mathbb{C}^2$ given in the symmetric gauge
\begin{align}
    R_{i,j}(u)=\frac{a(u)+b(u)}{2}\, \mathbbm{1}+ \frac{a(u)-b(u)}{2}\, \sigma^z_i\sigma^z_j+c(u)\, (\sigma^+_i\sigma^-_j+\sigma^-_i\sigma^+_j)\,,
\end{align}
where $\sigma^{\pm}=\frac{1}{2} (\sigma^{x}\pm\ri \sigma^{y})$ and $\sigma^{x,y,z}$ are the Pauli matrices. The weights depend on the spectral parameter $u\in \mathbb{C}$ and are given by 
\begin{align}
    a=\sinh(u+\ri \gamma)\,, \qquad b=\sinh(u)\,, \qquad c=\sinh(\ri \gamma)\,
\end{align}
and $\gamma\in \mathbb{R}$ parameterizes the anisotropy of the model. The R-matrix can be graphically depicted as in Figure \ref{Graphical Notation} and possesses the standard characteristics:
\begin{subequations}\label{R-properties}
\begin{equation}
   R^{t_it_j}_{i,j}(u)=R_{i,j}(u)\,,\quad   P_{i,j}R_{i,j}(u)P_{i,j}=R_{i,j}(u)\,,\quad   R_{i,j}(0)=\sinh (\ri\gamma ) P_{i,j} \,,\\
\end{equation}
as well as 
\begin{equation}
       R_{i,j}(u)R_{j,i}(-u)=\rho(u)\,\mathbbm{1}\,, \qquad R^{t_i}_{i,j}(u)R^{t_j}_{j,i}(-u-2\ri\gamma)=\rho(u+\ri\gamma)\,\mathbbm{1}\,.
\end{equation}
\end{subequations}
Here the superscript $t_i$ denotes the transposition in the associated space $V_i$, $P_{i,j}$ is the permutation matrix, and the scalar function $\rho(u)$ reads as
\begin{align}
    \rho(u)=\frac{1}{2}(\cos (2\gamma)-\cosh (2 u))\,.
\end{align}
\begin{figure}[t]
    \centering
   \begin{tikzpicture}
    \node[left] at (-1.5,0) {$R^{\gamma\delta}_{\alpha\beta}(u)=$}; 
    \node[above] at (0.25,0) {$u$}; 
    \node[left] at (-1,0) {$\alpha$}; 
    \draw[->,thick] (-1,0)--(1,0);
    \node[right] at (1,0) {$\gamma$}; 
    \node[below] at (0,-0.5) {$\beta$}; 
    \draw[->,thick] (0,-0.5)--(0,0.5);
    \node[above] at (0,0.5) {$\delta$}; 
    
    \draw[->,thick] (-6-1+11,0.0)--(-4-1+11,0.0);
    \node[left] at (-6-1+11,0.0) {$\alpha$};
    \node[right] at (-4-1+11,0.0) {$\beta\,,$};
    \cross{-5-1+11}{0};
    \node[left] at (-1.4-5-1+11,0.0) {$F^\beta_\alpha=$}; 
    \end{tikzpicture}
    \caption{The R-matrix and antidiagonal twist matrix $\sigma^x$ in graphical notation}
    \label{Graphical Notation}
\end{figure}
The R-matrix has, among others, the following symmetry property
\begin{equation}
    \begin{aligned}
    \kommu{R_{i,j}(u)}{\sigma^x_i\, \sigma^x_j}=0\,.
\end{aligned}
\end{equation}
This symmetry combined with the Yang-Baxter equation
\begin{align}
R_{j,k}(v)R_{i,k}(u)R_{i,j}(u-v)=R_{i,j}(u-v)R_{i,k}(u)R_{j,k}(v),  \label{YBE}
\end{align}
allows for the construction of a family of commuting operators in the following way
\begin{equation}
    \begin{aligned}
         \mathbbm{t}(u)=\tr_0\left(T_0(u)\,\sigma^x_0\right),&\qquad     \kommu{ \mathbbm{t}(u)}{ \mathbbm{t}(v)}=0\,,
\end{aligned}
\end{equation}
where the monodromy matrix $T_0(u)$, depending on the so-called inhomogeneities $\delta_i\in \mathbb{C}$, reads
\begin{align}
          T_0(u)=&R_{0,2L}(u- \delta_{2L})R_{0,2L-1}(u- \delta_{2L-1})\,...\,R_{0,1}(u- \delta_{1})\, .
\end{align}
The transfer matrix $\mathbbm{t}(u)$ acts on a circular lattice of $2L$ sites i.e. the Hilbert space $\mathscr{H}$  is given by $\mathscr{H}=V_1\otimes \dots \otimes V_{2L}$. The influence of the twist matrix $\sigma^x_0$ is encoded in the boundary conditions: given an operator $B_{j}$ acting non-trivially on the $j^{\text{th}}$ site we identify  
\begin{align}
    B_{2L+j}=\sigma^x_jB_{j}\sigma^x_j  \qquad j=1,\dots,2L\,. \label{AntiPeriodic_Condtion}
\end{align}
Note that these antidiagonal boundary conditions have a profound influence on the applicability of standard techniques used to diagonalise the transfer matrix. For example, the algebraic Bethe ansatz fails because a suitable pseudovacuum is unknown in the presence of $\sigma^x$ in $\mathbbm{t}(u)$.
Instead, one relies on the application of other methods e.g.\ the method of commuting transfer matrices by Baxter or Sklyanin's separation of variables \cite{Ba1972,Skly85,Skly92}. The former has been applied to the homogeneous six-vertex model with antidiagonal boundary conditions in \cite{YuBa95a,BBOY95}. It is straightforward to generalize this procedure to the inhomogeneous case. One arrives at the following TQ-equation
\begin{align}\label{dqeikjfi}
    \mathbbm{t}(u)\mathbbm{Q}(u)=\prod^{2L}_{j=1}\sinh(u-\delta_j+\ri\gamma) \mathbbm{Q}(u-\ri\gamma)-\prod^{2L}_{j=1}\sinh(u- \delta_j)\mathbbm{Q}(u+\ri\gamma)\,.
\end{align}
The Q-operator $\mathbbm{Q}(u)$ commutes with itself and the transfer matrix for different values of the spectral parameter $u$.  Its matrix elements are given in the appendix \ref{QQ} where we also present a truly explicit expression of $\mathbbm{Q}$  which, in contrast to the results of \cite{BBOY95}, does not contain any implicit operations such as matrix-inversion or matrix-multiplications. Hence, this form of the operator is particularly useful for numerical implementation.
\medskip

The above operator equation induces an equation for the eigenvalues $t(u), Q(u)$ of the transfer matrix and the Q-operator (see also \cite{NiWF09})
\begin{align}
    t(u)Q(u)=\prod^{2L}_{j=1}\sinh(u-\delta_j+\ri\gamma) Q(u-\ri\gamma)-\prod^{2L}_{j=1}\sinh(u-\delta_j)Q(u+\ri\gamma)\,. \label{TQ_eigenvalues}
\end{align}
This equation for $Q(u)$ can be solved by taking into account the analytic properties of the transfer matrix eigenvalues $t(u)$.  The transfer matrix is a Laurent polynomial in $\re^{u}$. The asymptotic behaviour of its eigenvalues regarding the spectral parameter $u$ is given by
\begin{align}\label{vunghfdj}
    \lim_{u\to \infty}t(u)e^{-2Lu}=O(e^{-u}).
\end{align}
Using \eqref{vunghfdj} one can deduce from \eqref{TQ_eigenvalues} that $Q(u)$ has to be of the form: 
\begin{align}\label{ddskdmk1}
    Q(u)=\prod^{2L}_{k=1}\sinh\left(\frac{1}{2}\left(u-v_k+\frac{\ri\gamma}{2}\right)\right)\,,
\end{align}
where the $v_k$ are unknown parameters, called the Bethe roots. The eigenvalue of the transfer matrix can be conveniently expressed in terms of the $v_k$
\begin{equation}
    \begin{aligned}
    t(u)=&\prod^{2L}_{j=1}\sinh(u-\delta_j+\ri\gamma)\prod^{2L}_{k=1}\frac{\sinh\left(\frac{1}{2}(u-v_k-\frac{\ri\gamma}{2})\right)}{\sinh\left(\frac{1}{2}(u-v_k+\frac{\ri\gamma}{2})\right)}\\&-\prod^{2L}_{j=1}\sinh(u-\delta_j)\prod^{2L}_{k=1}\frac{\sinh\left(\frac{1}{2}(u-v_k+\frac{3\ri\gamma}{2})\right)}{\sinh\left(\frac{1}{2}(u-v_k+\frac{\ri\gamma}{2})\right)}\,.
\end{aligned}
\end{equation}
Imposing analyticity on $t(u)$ at $u=v_k-\tfrac{\ri \gamma}{2}$  one obtains the Bethe equations for $v_k$: 
\begin{equation}\label{ddskdmk2}
    \begin{aligned}
   \prod^{2L}_{j=1} \frac{\sinh\left(v_k-\delta_j+\frac{\ri\gamma}{2}\right)}{\sinh\left(v_k-\delta_j-\frac{\ri\gamma}{2}\right)}=\prod^{2L}_{m=1}\frac{\sinh\left(\frac{1}{2}\left(v_k-v_m+\ri\gamma\right)\right)}{\sinh\left(\frac{1}{2}(v_k-v_m-\ri\gamma)\right)}\,,\qquad k=1,\dots,2L\,.
\end{aligned}
\end{equation}

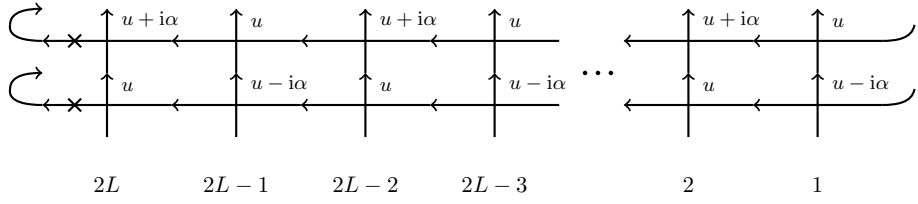
\begin{figure}
    \centering
    \begin{tikzpicture}[scale=0.85, transform shape]
\ecelll{0}{0}{$u$}
\ecelll{2}{0}{$u-\ri\alpha$}
\ecelll{4}{0}{$u$}
\ecelll{6}{0}{$u-\ri\alpha$}
\ecelll{9}{0}{$u$}
\ecelll{11}{0}{$u-\ri\alpha$}
\draw[-,thick,out=180,in=-100] (-1,0) to (-1.5,0.25);
\draw[->,thick,out=90,in=-180] (-1.5,0.25) to (-1,0.5);
\draw[-,thick,out=0,in=-100] (12,0) to (0.5+12,0.25);
\draw[-,thick,out=0,in=-100] (12,1) to (0.5+12,1.25);
\fill (7.4,0.5) circle (1pt);
\fill (7.6,0.5) circle (1pt);
\fill (7.8,0.5) circle (1pt);
\draw[-,thick,out=180,in=-100] (-1,1) to (-1.5,1.25);
\draw[->,thick,out=90,in=-180] (-1.5,1.25) to (-1,1.5);
\ecelll{0}{1}{$u+\ri\alpha$}
\ecelll{2}{1}{$u$}
\ecelll{4}{1}{$u+\ri\alpha$}
\ecelll{6}{1}{$u$}
\ecelll{9}{1}{$u+\ri\alpha$}
\ecelll{11}{1}{$u$}
\cross{-0.5}{0}
\cross{-0.5}{1}
\node[below] at (0,-1) {$2L$};
\node[below] at (2,-1) {$2L-1$};
\node[below] at (4,-1) {$2L-2$};
\node[below] at (6,-1) {$2L-3$};
\node[below] at (9,-1) {$2$};
\node[below] at (11,-1) {$1$};
\end{tikzpicture}
    \caption{Graphical representation of the product (\ref{Double_TM}) of two transfer matrices with staggering given by \eqref{Z_2_Staggering} by using the conventions defined in Figure~\ref{Graphical Notation}. }
    \label{Fig_Double_TM}
\end{figure}
In this work, we restrict ourselves to two-site periodically repeating inhomogeneities, parameterized by the so-called staggering parameter $\alpha$: 
\begin{align}
    \delta_{2j}=-\frac{\ri\alpha}{2},\qquad \delta_{2j-1}=\frac{\ri\alpha}{2} \qquad \text{with } \quad j=1,\dots, L \label{Z_2_Staggering}\,.
\end{align}
In addition to this horizontal staggering, we also introduce a vertical staggering by multiplying two transfer matrices with shifted arguments: 
\begin{align}
    \mathbbm{T}(u)=\mathbbm{t}\left(u+\frac{\ri\alpha}{2}\right)\mathbbm{t}\left(u-\frac{\ri\alpha}{2}\right)\,. \label{Double_TM}
\end{align}
This two-row transfer matrix can be graphically depicted in Figure \ref{Fig_Double_TM}.  It reduces to the two-site translation operator when evaluated at $u=0$. Therefore, we obtain a local Hamiltonian by taking the logarithmic derivative 
\begin{align}
\label{eq:hfromt}
   \mathbbm{H}= -\ri\left.\deri{u}\log{ \mathbbm{T}(u)}\right|_{u=0}-L \left(\cot (\alpha -\gamma )-\cot (\alpha +\gamma )-2 \cot (\gamma )\right)\,.
\end{align}
In terms of the Pauli matrices subject to the boundary conditions (\ref{AntiPeriodic_Condtion}), the Hamiltonian reads
\begin{equation}
\label{End_H-Op}
\begin{aligned}
    \mathbbm{H}=&-\frac{1}{2\sin(\gamma)\rho(\ri\alpha)}\bigg\{ -L\cos (\gamma )(1+\cos(2\alpha)-2\cos(2\gamma)) \\
    &-2\sin^2(\gamma)\sum^{2L}_{j=1}\cos(\gamma) \sigma^{z}_{j}\sigma^{z}_{j+1}+2\cos(\alpha)(\sigma^{+}_{j}\sigma^{-}_{j+1}+\sigma^{-}_{j}\sigma^{+}_{j+1})\\ 
\qquad&+\cos(\gamma)\sin^2(\alpha) \sum^{2L}_{j=1} \sigma^{z}_{j}\sigma^{z}_{j+2}+2(\sigma^{+}_{j}\sigma^{-}_{j+2}+\sigma^{-}_{j}\sigma^{+}_{j+2})\\
\qquad &+\sin(\alpha)\sin(2\gamma)\sum^{2L}_{j=1}(-1)^{j+1}\sigma^z_j\sigma^+_{j+1}\sigma^-_{j+2}+(-1)^{j}\sigma^z_j\sigma^-_{j+1}\sigma^+_{j+2}\\
&+\sin(\alpha)\sin(2\gamma)\sum^{2L}_{j=1}(-1)^{j+1}\sigma^+_j\sigma^-_{j+1}\sigma^z_{j+2}+(-1)^{j}\sigma^-_j\sigma^+_{j+1}\sigma^z_{j+2}\\
&+\sin(\gamma)\sin(2\alpha)\sum^{2L}_{j=1}(-1)^{j+1}\sigma^-_j\sigma^z_{j+1}\sigma^+_{j+2} +(-1)^j\sigma^+_j\sigma^z_{j+1}\sigma^-_{j+2}
 \bigg\}\,\\
&-L\cot(\alpha-\gamma)+2L\cot(\gamma)+L\cot(\alpha+\gamma)\,.
\end{aligned}
\end{equation}

It should be pointed out that the Hamiltonian is not self-adjoint for generic values of the anisotropy and the staggering. Moreover, the global $O(2)$ symmetry of the model is broken to two discrete $\mathbb{Z}_2$ symmetries in the antidiagonal case considered here, see Ref.~\cite{ABGR88}. They can be expressed as products over Pauli matrices
\begin{align}
    \mathbbm{G}=\prod^{2L}_{j=1} \sigma^z_j\,,\qquad \mathbbm{C}=\prod^{2L}_{j=1} \sigma^x_j.\label{afjghnee}
\end{align} 

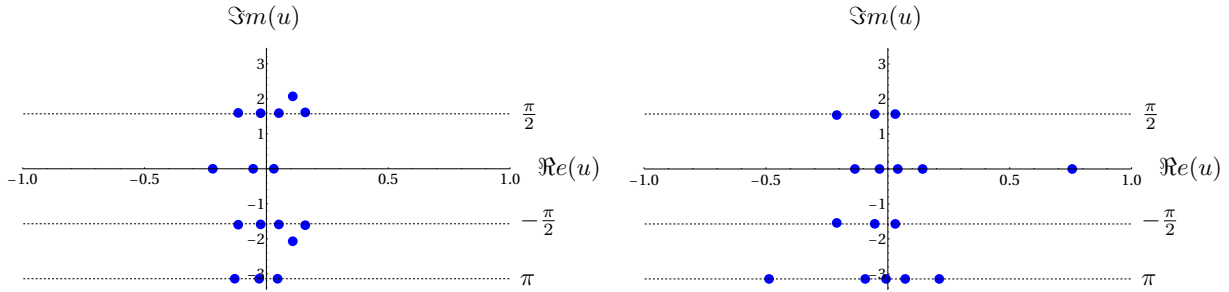
\begin{figure}[t]
    \centering
    \begin{minipage}[b]{.49\linewidth}
    \scalebox{0.93}{
\begin{tikzpicture}
\node at (0,0) {\includegraphics[width=0.9\linewidth]{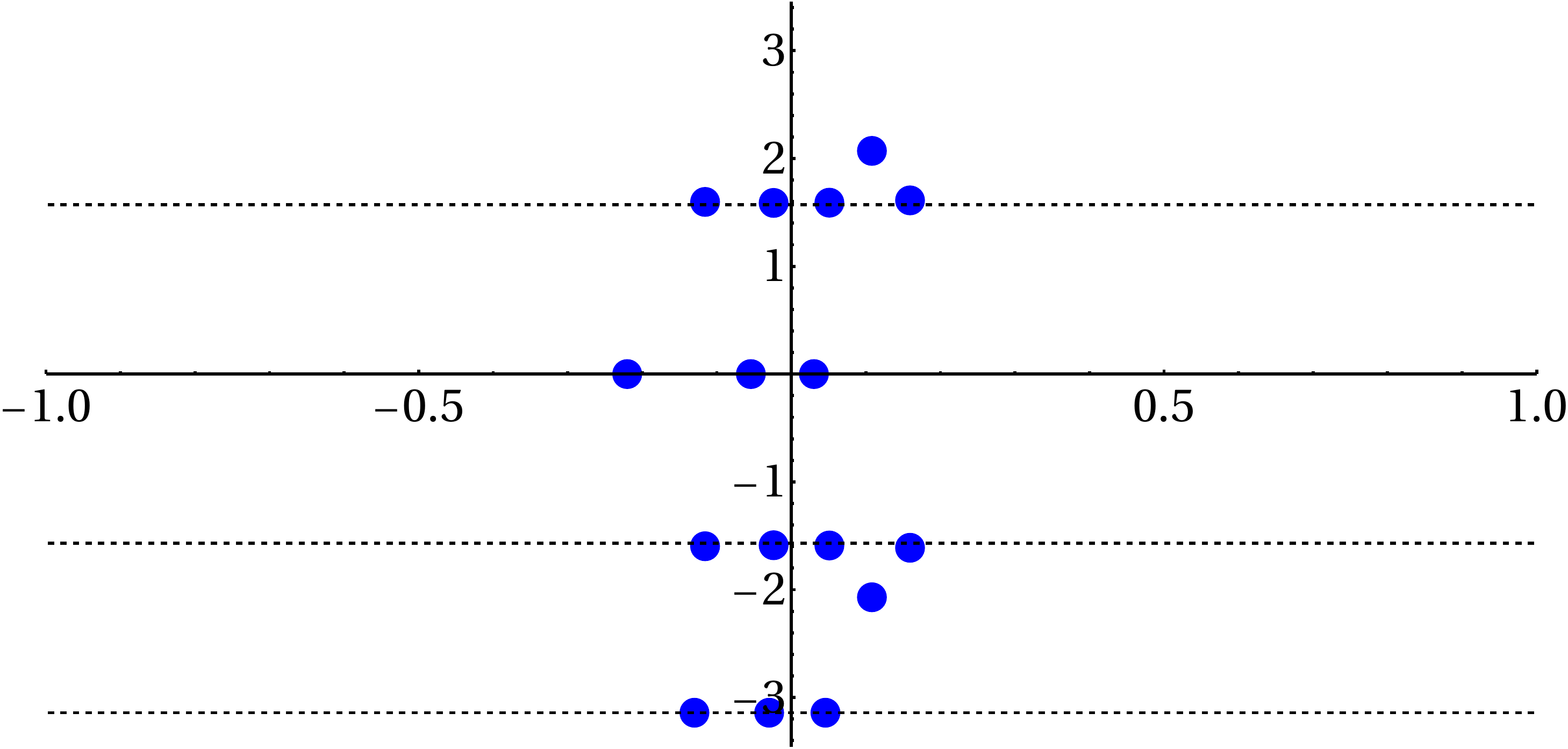}};
\node at (0.00,2.15)   {\small$\Im m(u)$};
\node[anchor=west] at  (3.75,0.00)  {\small$\Re e(u)$};
\node[anchor=west] at  (3.5,0.750)  {\small$\tfrac{\pi}{2}$};
\node[anchor=west] at  (3.5,-0.7650)  {\small$-\tfrac{\pi}{2}$};
\node[anchor=west] at  (3.5,-1.575)  {\small$\pi$};
\end{tikzpicture}
}
 \end{minipage}
 \begin{minipage}[b]{.49\linewidth}
      \scalebox{0.93}{
\begin{tikzpicture}
\node at (0,0) {\includegraphics[width=0.9\linewidth]{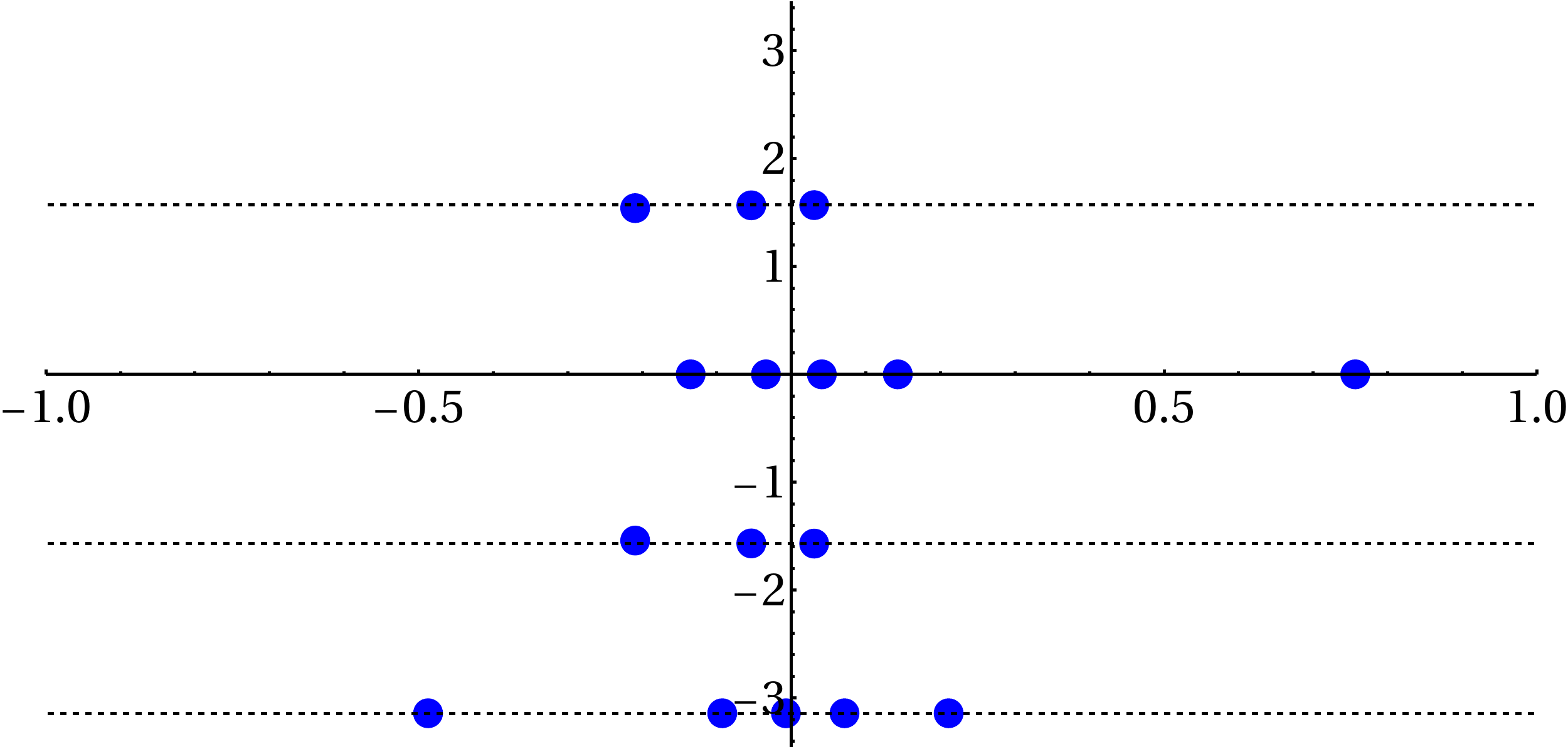}};
\node at (0.00,2.15)   {\small$\Im m(u)$};
\node[anchor=west] at  (3.75,0.00)  {\small$\Re e(u)$};
\node[anchor=west] at  (3.5,0.750)  {\small$\tfrac{\pi}{2}$};
\node[anchor=west] at  (3.5,-0.7650)  {\small$-\tfrac{\pi}{2}$};
\node[anchor=west] at  (3.5,-1.575)  {\small$\pi$};
\end{tikzpicture}
}
\end{minipage}
    \caption{The different Bethe root configurations for an excited state related by the transformation (\ref{eq:dual2}) for $2L=16$, $\gamma=2\pi/5$, $\alpha=\pi/2$: in both cases the roots tend to align on lines with constant imaginary part.  The deviations from these lines, however, are smaller for the dual root configuration (right panel) which makes this parameterization more convenient for the finite size analysis (see \cite{DATA} for the numerical data used in this  figure). 
    }
    \label{Dual_Vs_Non_Dual}
\end{figure}

In addition, the Hamiltonian possesses duality transformations \cite{IkJS08,FrMa12,BKKL21b} relating models with different values of $\gamma$ and $\alpha$: under the action of 
\begin{align}
    \mathcalboondox{D}=\prod^{L}_{i=1} P_{2i-1,2i}R_{2i-1,2i}(\ri \alpha)
\end{align}
the model is mapped to a different staggering parameter
\begin{align}
   \mathcalboondox{D}\,\left.\mathbbm{H}\right|_{\alpha}\, \mathcalboondox{D}^{-1}=\left. \mathbbm{H}\right|_{\alpha\to \pi-\alpha}\,. \label{ntjfn}
\end{align}
Further, under the transformation
\begin{align}
\label{eq:dual2}
    \mathcal{D}:\quad\begin{cases} \alpha \to \pi-\alpha\\
    \gamma\to \pi-\gamma\end{cases}
\end{align}
the high energy and low energy spectra are interchanged 
\begin{align}
    \mathcal{D}(\mathbbm{H})=-\mathbbm{H}\,.
\end{align}
By negation,  we can neutralise the influence of the duality transformation $\mathcal{D}$ on the Hamiltonian. The resulting Hamiltonian (and its spectrum) is then invariant. In contrast, the eigenvalues of the transfer matrix and the Q-operator are periodic functions in $\gamma$ but with period $4\pi$ (e.g. see \eqref{ddskdmk1}). Hence, they are modified under the duality transformation.
This leads to the fact that there exist \emph{two} interchangeable sets of Bethe roots describing the spectrum of $\mathbbm{H}$; see for comparison the Figure \ref{Dual_Vs_Non_Dual}. 
For technical reasons, the Bethe root patterns of the duality transformed roots are sometimes slightly more convenient for further study. Therefore, we always use the `dual' root configurations in the following.  

Besides the family of commuting operators originating from the \emph{product} of individual transfer matrices in (\ref{Double_TM}), one can also study the family generated by the \emph{quotient} of these transfer matrices: 
\begin{align}
    \mathbbm{K}(u)=\frac{\mathbbm{t}\left(u-\frac{\ri\alpha}{2}\right)}{\mathbbm{t}\left(u+\frac{\ri\alpha}{2}\right)}\,. \label{Gen_QI}
\end{align}
The logarithm of  (\ref{Gen_QI}) at $u=0$
\begin{align}
    \mathbbm{B}=\log\left(\mathbbm{K}(0)\right)\label{QI}
\end{align}
is called the quasi-momentum operator. It plays an important role in identifying the scaling limit of the staggered six-vertex model for the boundary conditions studied in previous works \cite{IkJS12,BKKL21a,BKKL21b,FrGe22,RPJS20,FrGe23,FrGK23}. 

In terms of the Bethe roots the eigenenergies and the eigenvalues of the quasi-momentum operator can be expressed as
\begin{align}
    E    &=\sum^{2L}\label{uhu}_{k=1}\frac{\sin\left(\frac{\gamma}{2}\right)}{\cosh(v_k-\frac{\ri\alpha}{2})-\cos(\frac{\gamma}{2})}+\frac{\sin\left(\frac{\gamma}{2}\right)}{\cosh(v_k+\frac{\ri\alpha}{2})-\cos(\frac{\gamma}{2})}\,,\\
     B&= \log \left( \frac{\sin(\gamma+\alpha)}{\sin(\gamma-\alpha)} \right)^L+\sum^{2L}_{k=1}\log\left( \frac{\cosh(v_k)-\cos\left(\frac{\alpha-\gamma}{2} \right)
    }{\cosh(v_k)-\cos\left(\frac{\alpha+\gamma}{2}\right) } \right)\,.\label{oxnsh}
\end{align}

\section{Finite-size studies}
\label{sec:fs}
\begin{figure}[t]
    \centering
    \begin{minipage}[b]{.49\linewidth}
    \scalebox{0.93}{
\begin{tikzpicture}
\node at (0,0) {\includegraphics[width=0.9\linewidth]{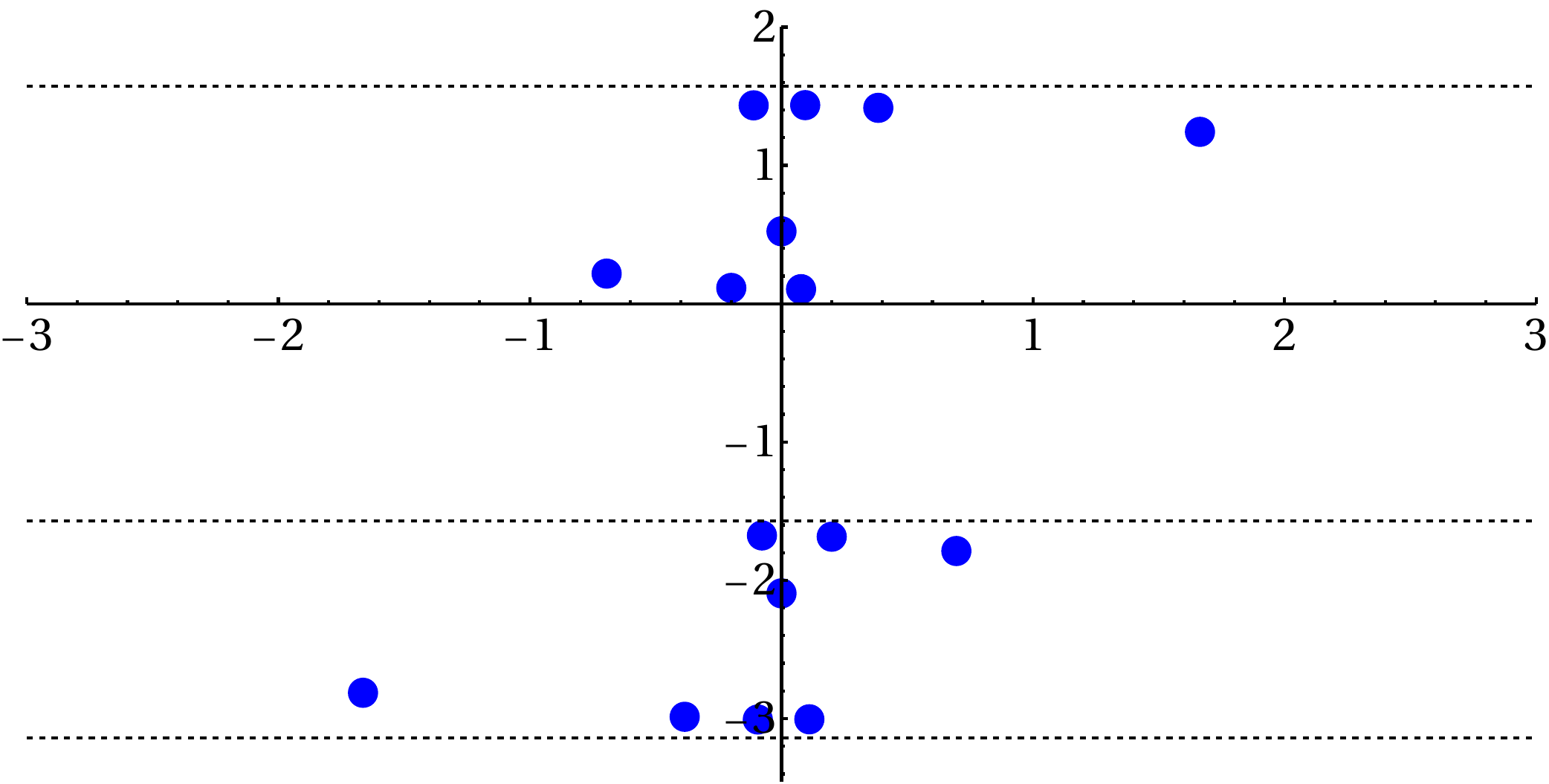}};
\node at (0.00,2.15)   {\small$\Im m(u)$};
\node[anchor=west] at  (3.75,0.45)  {\small$\Re e(u)$};
\node[anchor=west] at (3.6,1.4)   {\small$\tfrac{\pi}{2}$};
\node[anchor=west] at (3.6,-0.6)   {\small$-\tfrac{\pi}{2}$};
\node[anchor=west] at (3.6,-1.6)  {\small$-\pi$};
\end{tikzpicture}
}
 \end{minipage}
 \begin{minipage}[b]{.49\linewidth}
      \scalebox{0.93}{
\begin{tikzpicture}
\node at (0,0) {\includegraphics[width=0.9\linewidth]{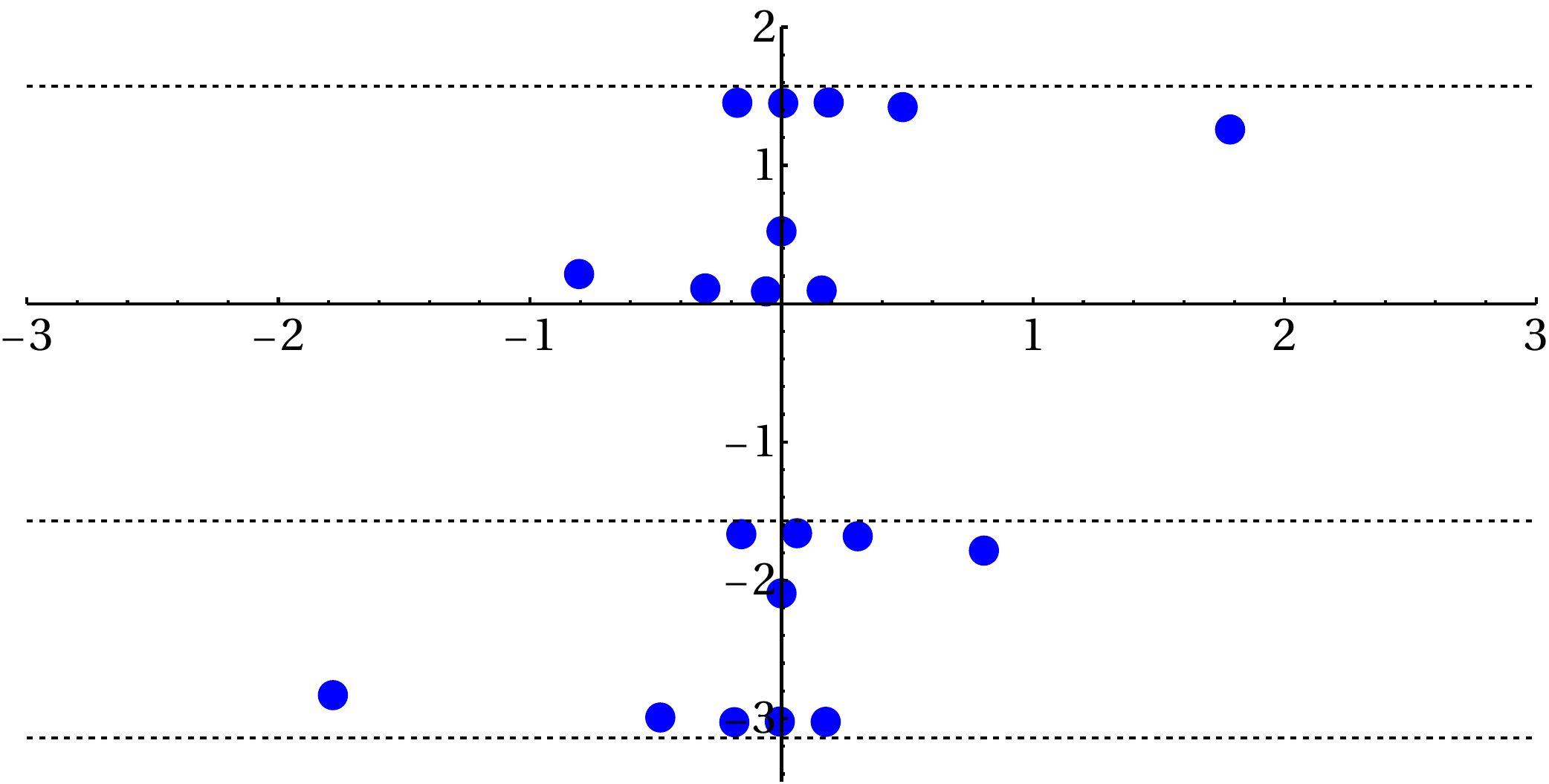}};
\node at (0.00,2.15)   {\small$\Im m(u)$};
\node[anchor=west] at  (3.75,0.45)  {\small$\Re e(u)$};
\node[anchor=west] at (3.6,1.4)   {\small$\tfrac{\pi}{2}$};
\node[anchor=west] at (3.6,-0.6)   {\small$-\tfrac{\pi}{2}$};
\node[anchor=west] at (3.6,-1.6)  {\small$-\pi$};
\end{tikzpicture}
}
\end{minipage}
    \caption{Construction of an  RG-trajectory: shown are the dual Bethe root configurations of particular states for $\gamma=\frac{\pi}{6}$, $\alpha=\frac{\pi}{2}$ for $2L=16$ (left panel) and $20$ (right panel): the states are identified as members of the  same RG-trajectory since the root patterns are characterized by their main features, i.e.\
    two purely imaginary roots at $\frac{\ri \pi}{6}$ and $-\frac{2\ri\pi}{3}$ apart from the bulk of roots with imaginary parts near $0,\pm\frac{\pi}{2},-\pi$.
    }
    \label{RG_Construction}
\end{figure}
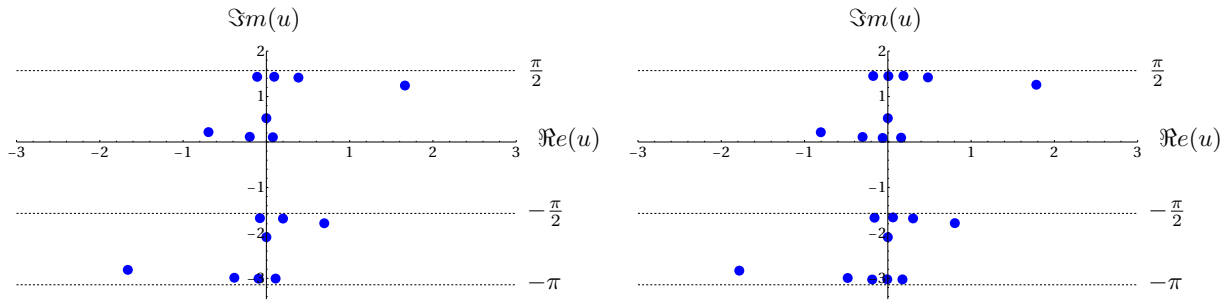

In the following sections, we investigate the low energy spectrum of the lattice model for increasing system sizes to obtain information about the effective field theory arising in its scaling limit. For the case at hand, an integrable lattice model, this task is facilitated by the description of individual states in terms of their corresponding Bethe root configurations $\{v_k\}^{2L}$: 
starting from a particular eigenstate $\ket{\Psi_{L_{\rm in}}}$ of the model with $L_{\rm in}$ sites we construct so-called RG-trajectories $\{\ket\Psi_L\}$ by grouping states $\ket{\Psi_{L}}$ with $L=L_{\rm in},L_{\rm in}+2,L_{\rm in}+4,\dots$ with Bethe roots forming qualitatively similar patterns, see Figure \ref{RG_Construction} for an illustration.
%
This allows to construct RG-trajectories $\{\ket{\Psi_{L}}\}$ for system sizes up to $L \sim 2000$ by merely solving the Bethe ansatz equation without relying on a direct diagonalization of the Hamiltonian. The latter is an impossible task as the size of the Hilbert space grows exponentially with the system size $L$. 

The  initial data for this construction, i.e.\ the Bethe root configurations for some small system size $L_{\rm in}$, 
are obtained by a simultaneous diagonalization of the Hamiltonian and the Q-operator.  The root configurations for the few hundred lowest energy states are obtained by finding the zeros of $Q(u)$, see \eqref{ddskdmk1}.  These are then taken as the seed for the construction of the corresponding RG trajectories via the procedure described above.  
\medskip

Using this method, we can study the finite-size scaling of the energies. For a closed spin chain at its critical point, we expect  that the field theory is conformal invariant \cite{Polyakov1970} and that the energies follow the asymptotic behaviour for large system sizes $L$ \cite{Cardy86a,BlCN86}
\begin{align}
\label{Xeff}
   E\asymp  Le_\infty+ \frac{2\pi v_{\rm F}}{L}X_{{\rm eff}}\,, \qquad 
    X_{{\rm eff}}=-\frac{c}{12}+ \Delta+ \bar{\Delta} +{\tt L}+ \bar{{\tt L}}
\end{align}
where the the constant $e_\infty$ is the bulk energy density, $v_{\rm F}$ the Fermi velocity and $X_{{\rm eff}}$ is called the effective scaling dimension. The latter is given in terms of the universal central charge $c$ of the underlying CFT and the conformal weights $\Delta$, $\bar{\Delta}$ of the conformal primaries while ${\tt L}$ and $ \bar{{\tt L}}$ denote the level of the descendant fields. 

In the next three sections, we study all parametric regimes given by
\begin{itemize}
    \item Phase I:  $\quad\,\,\alpha<\min(\gamma,\pi-\gamma)\qquad$ (or, by duality (\ref{ntjfn}), $\alpha>\max(\gamma,\pi-\gamma) $).
    \item Phase II:  $\quad\gamma>\pi/2$ and $\pi-\gamma<\alpha<\gamma$.
    \item Phase III:  $\quad\gamma<\pi/2$ and $\gamma<\alpha<\pi-\gamma$.
\end{itemize}

\subsection{Phase I}
In this phase, the roots for the ground state are arranged in a simple pattern given by 
\begin{align}
    v^{(1)}_k=x_k\,,\qquad k=1,\dots, L\qquad \qquad v^{(2)}_j=y_j-\ri \pi \,,\qquad j=1,\dots, L\label{ghvnpgnf}
\end{align}
In terms of the real parts $x_k,y_j$, the logarithmic Bethe equations for the ground state read
\begin{align}
    c^x(x_k)=2\pi I^x_m,\qquad c^y(y_j)=2\pi I^y_j.
\end{align}
where the counting functions are defined as 
\begin{align}
    c^x(x)&=-L\phi\left(x,\frac{\gamma+\alpha}{2}\right)- L\phi\left(x,\frac{\gamma-\alpha}{2} \right)+\sum^{L}_{k=1} \phi\left(\frac{1}{2}(x-x_k),\frac{\gamma}{2}\right)-\sum^{L}_{k=1} \psi\left(\frac{1}{2}(x-y_k),\frac{\gamma}{2}\right)\,,\\
    c^y(x)&=-L\phi\left(x,\frac{\gamma+\alpha}{2}\right)- L\phi\left(x,\frac{\gamma-\alpha}{2} \right)-\sum^{L}_{k=1} \psi\left(\frac{1}{2}(x-x_k),\frac{\gamma}{2}\right)+\sum^{L}_{k=1} \phi\left(\frac{1}{2}(x-y_k),\frac{\gamma}{2}\right)\,,
\end{align}
and we have introduced the functions
\begin{align}\phi(x,y)=2\arctan\left(\tanh(x)\cot(y)\right)\,, \qquad\psi(x,y)=2\arctan\left(\tanh(x)\tan(y)\right)\,.
\end{align}
The  Bethe integers $I^{x,y}_k$ take (half-)integer values for (odd) even $L$. Within the root density approach \cite{YaYa69}, we obtain the Fermi velocity
\begin{align}
    v_{\rm F}=\frac{\pi}{(\pi-\gamma)}
\end{align}
and the ground state energy density in the thermodynamic limit
\begin{align*}
    e_\infty=& \frac{2\sin(\pi-\gamma)\cos\left( \frac{\pi\alpha}{2(\pi-\gamma)} \right)}{(\pi-\gamma)}\, \int^{\infty}_{-\infty} \, {\rm d}x\, \frac{(\cos((\pi-\gamma))-\cos(\alpha)\cosh(2x))}{\sinh\left(x-\frac{\ri (\alpha-(\pi-\gamma))}{2}\right)\sinh\left(x+\frac{\ri (\alpha-(\pi-\gamma))}{2}\right)}\\
    &\times \frac{1}{\sinh\left(x-\frac{\ri (\alpha+(\pi-\gamma))}{2}\right)\sinh\left(x+\frac{\ri (\alpha+(\pi-\gamma))}{2}\right)} \frac{\cosh(\frac{\pi x }{(\pi-\gamma)})}{\cosh\left(\frac{2\pi x}{(\pi-\gamma)}\right)+\cos\left( \frac{\pi \alpha}{(\pi-\gamma)}\right)}.
\end{align*}
For $\alpha= 0$ this reduces to the known results of the homogeneous case obtained in \cite{NiWF09}, if one identifies $\gamma\mapsto \pi-\gamma$ and takes into account that the number of unit cells is doubled in the homogenous limit.

The majority of the Bethe roots of a low energy excitation above the ground state is still given by \eqref{ghvnpgnf} while some roots are located in the complex plane either as strings or single complex roots.  Our numerical work shows that the effective scaling dimensions are given by
\begin{align}
    X_{\rm eff}=-\frac{1}{24}+\frac{k}{2} \qquad \qquad k=0,1,2,\dots \,.\label{ghekn}
\end{align}
Note that, the anisotropy $\gamma$ and the staggering parameter $\alpha$ do not influence the effective scaling dimensions. Further, the degeneracy of each scaling dimension \eqref{ghekn} is given by 
\begin{equation}\label{ghnlc}
    \begin{aligned}
    \sum^{\infty}_{k=0} {\tt deg}(k)\,{\tt q}^{\frac{1}{24}+\frac{k}{2}} &=2\,{\tt q}^{\frac{1}{24}}\prod^{\infty}_{m=0}\left(1-{\tt q}^{m+\frac{1}{2}}\right)^{-2}\\
    &=2\left({\tt q}^{\frac{1}{24}}+2{\tt q}^{\frac{13}{24}}+3{\tt q}^{\frac{25}{24}}+6{\tt q}^{\frac{37}{24}}+\dots\right)
\end{aligned}
\end{equation}
We have checked this up to the indicated order numerically. The degeneracy formula \eqref{ghnlc} corresponds to twice the partition function of a CFT built from the (unique) irreducible representation of a twisted $U(1)$ Kac-Moody algebra, one for each sector of the $\mathbb{Z}_2$-symmetry $\mathbb{G}$ (\ref{afjghnee}).
The central charge is one ($c=1$) and the zero mode of the Virasoro algebra takes the form \cite{GeRS88}
\begin{align}
    L_0= \sum_{\mu\in\mathbb{N}+\frac{1}{2}} \, \mu \, a^\dagger_\mu \, a_\mu+\frac{1}{16}\,,
\end{align}
where the operators $a_m$ with $m\in \mathbb{Z}+\tfrac{1}{2}$ form a Heisenberg algebra: 
\begin{align}\left[a_k,a^\dagger_p\right]= k \, \delta_{k,p} \,.
\end{align}
The space of states for a given eigenvalue $g\in\{\pm\}$ of $\mathbbm{G}$ can then be described as a Fock space originating from a unique vacuum with weight $\frac{1}{16}$. Taking the antichiral modes into account, the degeneracy of each Fock space level is given by  \eqref{ghnlc}. Alternatively, the Fock space can be decomposed into a direct sum of representations of the Virasoro algebra, see Ref.~\cite{GeRS88}.
It should be stressed, that this CFT has been shown to describe the universal behaviour of the homogeneous model \cite{ABGR88}. Hence, the staggering is an irrelevant deformation of the homogenous model in this phase.

In the homogeneous model the operator content of the underlying CFT can be grouped into sectors with the same transformation behaviour under the two $\mathbb{Z}_2$ symmetries of the lattice model, see Eq.~(16) of Ref.~\cite{ABGR88}. The same holds for the case at hand.

\subsection{Phase II}
\begin{figure}[t]
    \centering
    \begin{minipage}[b]{.49\linewidth}
    \scalebox{0.93}{
\begin{tikzpicture}
\node at (0,0) {\includegraphics[width=0.9\linewidth]{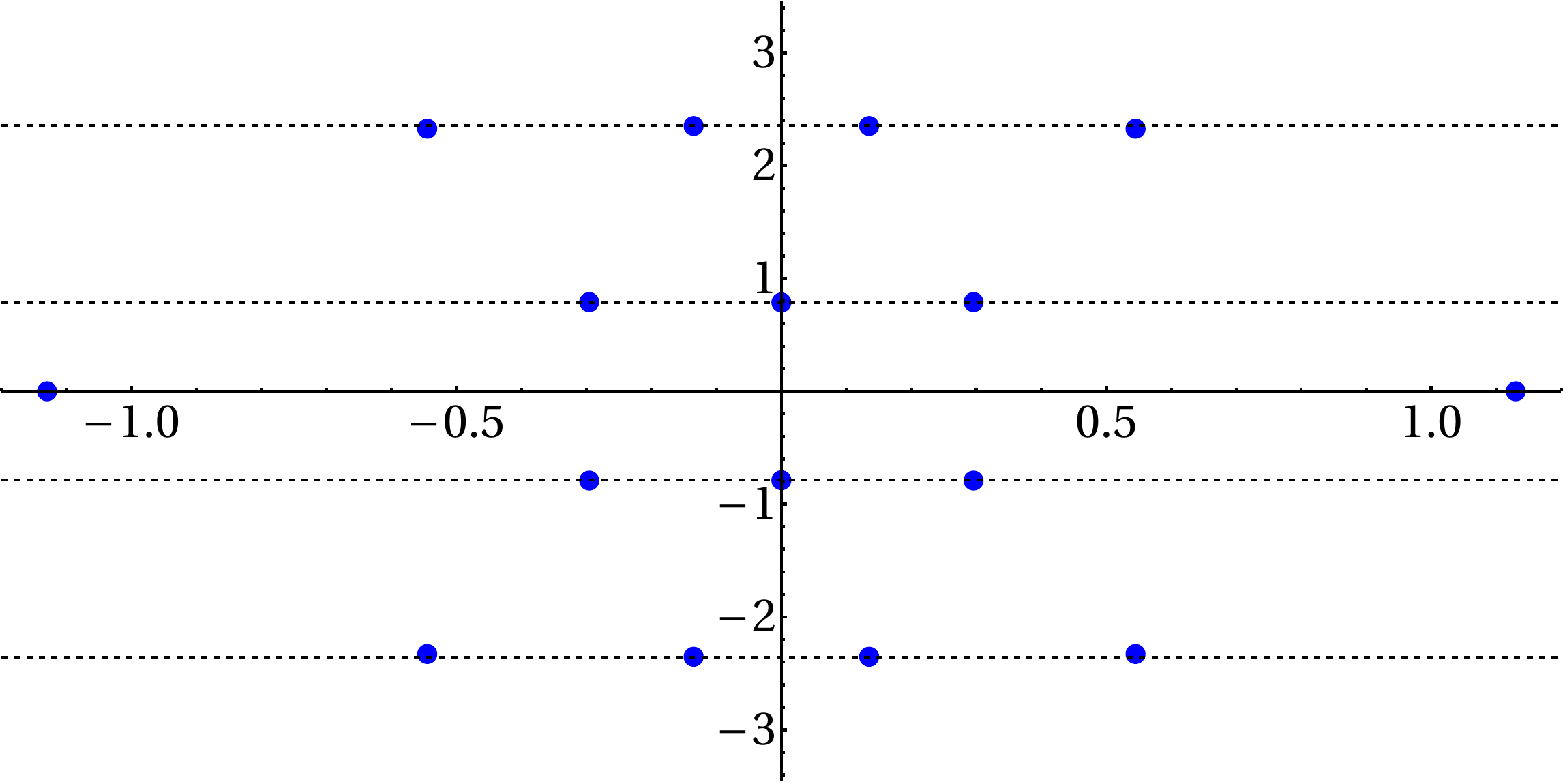}};
\node at (0.00,2.15)   {\small$\Im m(u)$};
\node[anchor=west] at  (3.75,0.00)  {\small$\Re e(u)$};

\node[anchor=west] at (3.6,1.275)   {\small$\tfrac{\pi+\alpha}{2}$};
\node[anchor=west] at (3.6,0.42)   {\small$\tfrac{\pi-\alpha}{2}$};
\node[anchor=west] at (3.6,-0.42)   {\small$\tfrac{\alpha-\pi}{2}$};
\node[anchor=west] at (3.6,-1.225)  {\small$-\tfrac{\pi+\alpha}{2}$};
\end{tikzpicture}
}
 \end{minipage}
 \begin{minipage}[b]{.49\linewidth}
      \scalebox{0.93}{
\begin{tikzpicture}
\node at (0,0) {\includegraphics[width=0.9\linewidth]{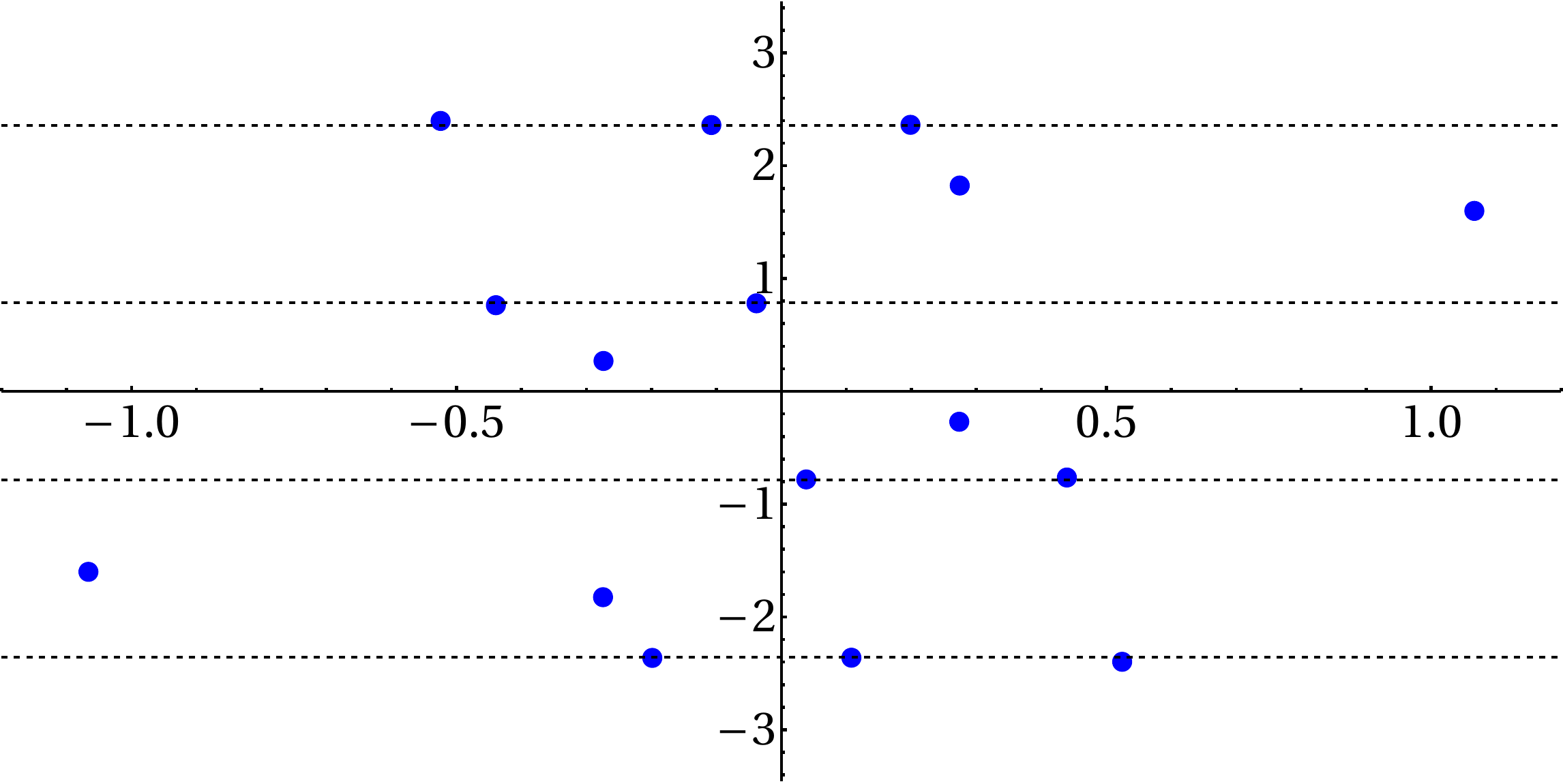}};
\node at (0.00,2.15)   {\small$\Im m(u)$};
\node[anchor=west] at  (3.75,0.00)  {\small$\Re e(u)$};

\node[anchor=west] at (3.6,1.275)   {\small$\tfrac{\pi+\alpha}{2}$};
\node[anchor=west] at (3.6,0.42)   {\small$\tfrac{\pi-\alpha}{2}$};
\node[anchor=west] at (3.6,-0.42)   {\small$\tfrac{\alpha-\pi}{2}$};
\node[anchor=west] at (3.6,-1.225)  {\small$-\tfrac{\pi+\alpha}{2}$};
\end{tikzpicture}
}
\end{minipage}
    \caption{Left (right) plot displays the dual Bethe-root configuration of the ground state (excited state) for $\gamma=\frac{2\pi}{3}$ and $\alpha=\tfrac{\pi}{2}$ in phase II in the complex $u$-plane for $2L=16$.}
    \label{BR_GS_Ex_A_2}
\end{figure}
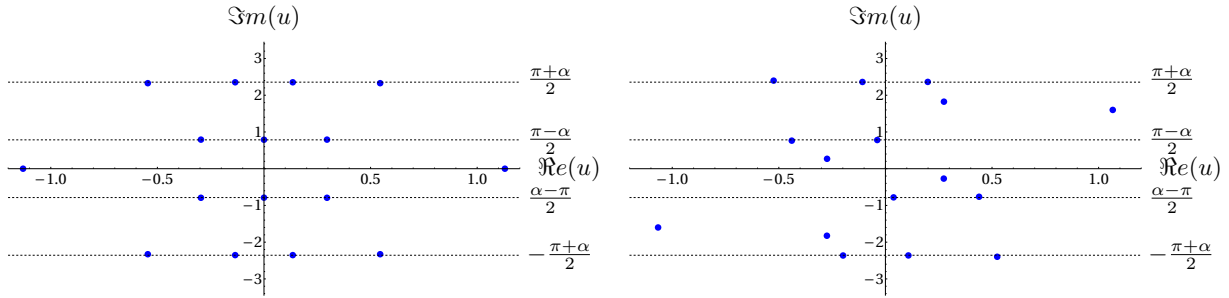

In this phase, we find that the Bethe roots parameterizing the ground state for even $L$ take values
\begin{align}
\label{eq:stringsII}
    v^{(1)}_k=x_k\pm\frac{\ri(\pi+\alpha)}{2}+\ri \varepsilon^{(1)}_k,\qquad v^{(2)}_k=y_k\pm\frac{\ri(\pi-\alpha)}{2}+\ri \varepsilon^{(2)}_k \qquad k=1,\dots,L-1
\end{align}
plus two additional roots on the real axis. The deviations $\varepsilon^{(1,2)}_k\in \mathbb{R}$ tend to zero as $L\to \infty$. Excitations above the ground state are generated by removing roots from the lines of the ground state pattern and  placing them in the complex plane either as single roots or as complexes such as strings.  The Bethe root configuration of the ground state and for one excited state is depicted in Figure \ref{BR_GS_Ex_A_2}.

Using the regular pattern (\ref{eq:stringsII}) of the ground state Bethe roots, we find that the energy density obtained by means of the standard root density approach coincides with that for (quasi)-periodic boundary conditions, i.e.\ \cite{KoLu2021}
\begin{align}
    e_\infty=-\frac{\sin (\gamma )}{ \pi - \gamma }\int^{\infty}_{-\infty}\mathrm{d}x\, \frac{1}{\cosh(\frac{\pi x}{2\pi-2\gamma})} \left( \frac{1}{\cosh(x)+\cos(\gamma)}+\frac{1}{\cosh(x-2\ri \alpha)+\cos(\gamma)} \right).
\end{align}
The Fermi velocity is given by 
\begin{align}
        v_{\rm F}=\frac{\pi }{\pi -\gamma }\,.
\end{align}
Our finite-size analysis of the spectrum results in the following spectrum of effective scaling dimensions
\begin{align}
    X_{{\rm eff}}(k)=\frac{1}{12}+\frac{k}{2}, \qquad k=0,1,2, \dots
\end{align}
-- again independent of the anisotropy $\gamma$ and the  staggering parameter $\alpha$.  Here the degeneracies ${\tt deg}(k)$ of $X_{\textrm{eff}}(k)$ are given by the generating function 
\begin{equation}
\label{Gen_Deg}
    \begin{aligned}
    \sum^{\infty}_{k=0} {\tt deg}(k)\,{\tt q}^{\frac{1}{12}+\frac{k}{2}} &=4\,{\tt q}^{\frac{1}{12}}\prod^{\infty}_{m=0}\left(1-{\tt q}^{m+\frac{1}{2}}\right)^{-4}\\
    &=4\left({\tt q}^{\frac{1}{12}}+4{\tt q}^{\frac{7}{12}}+10{\tt q}^{\frac{13}{12}}+24{\tt q}^{\frac{23}{12}}+\dots\right)
\end{aligned}
\end{equation}
We have checked this up to the indicated order numerically. Note that this is the square of the corresponding function (\ref{ghnlc}) in phase I.  As a result we propose that the CFT describing the scaling limit of the staggered vertex model in phase II has a central charge $c=2$ and is built from two copies of the twisted $U(1)$ Kac-Moody algebra.
Here, the zero mode of the Virasoro algebra takes the form
\begin{align}
\label{eq:L0}
    L_0= \sum_{\mu\in\mathbb{N}+\frac{1}{2}} \, \mu \, a^\dagger_\mu \, a_\mu+\sum_{\nu\in\mathbb{N}+\frac{1}{2}} \, \nu \,  b^\dagger_\nu \, b_\nu +\frac{2}{16}\,,
\end{align}
where the operators $a_m$, $b_m$ with $m\in \mathbb{Z}+\tfrac{1}{2}$ form two independent Heisenberg algebras. In fact, all operators commute among each other except for 
\begin{align}\left[a_k,a^\dagger_p\right]=\left[b_k,b^\dagger_p\right]= k \, \delta_{k,p} \,.
\end{align}
We note that according to (\ref{Gen_Deg}) the vacuum of the bosonic Fock space is fourfold degenerate. In the lattice model this multiplicity arises from (\ref{ntjfn}), but only on the self-dual line, $\alpha=\pi/2$.  For other values of the staggering the $\mathbb{Z}_2$ symmetries (\ref{afjghnee}) imply a doubly degenerate ground state, as in phase I.  This indicates that an additional $\mathbb{Z}_2$ symmetry emerges in the scaling limit of the lattice model throughout phase II and that the staggering parameter $\alpha$ is an irrelevant deformation of the model in phase II, similar to the periodic model and as in phase I.

For periodic boundary conditions, the scaling limit of the staggered model in this phase was shown to be related to one compact boson and two Majorana fermions \cite{IkJS10,KoLu2021}. Note that the present findings for antidiagonal BC do not contradict the results of the staggered model with periodic ones.  This is due to the fact that the modules of the Majoranas will start to interfere with the ones of the boson once the antiperiodicity is imposed. This can lead to a restructuring of all the modules, which will then be equivalent to the case at hand.  

\subsection{Phase III}
Bulk quantities, such as the energy density of a spin chain in the thermodynamic limit, do not depend on the specific choice of boundary conditions imposed. We have already seen this phenomenon in phases I and II. Hence, we expect that in phase III, the energy density will be identical to the known one of the periodic model \cite{FrSe14}:
\begin{align}
        e_\infty&=-2\int^{\infty}_{-\infty}\text{d}\omega  \frac{\sinh(\frac{\gamma \omega}{2})\left(\sinh \left(\frac{\pi\omega}{2}-\frac{\omega \gamma}{2}   \right)\cosh(\frac{\omega\pi}{2}-\alpha \omega)-\sinh(\frac{\gamma \omega}{2})\right)}{\sinh(\frac{\omega \pi}{2})\sinh((\frac{\pi-2\gamma}{2})\omega)}\,.\label{eInf}
\end{align}

To obtain the above energy density, we find that the Bethe roots of the low-lying energy excitation should align on the following four lines in the thermodynamic limit 
\begin{equation}\label{Bethe_Hypothesis}
\begin{aligned}
    v^{(1)}_k=x_k\,,\qquad \qquad 
    v^{(2)}_k=y_k-\ri \pi\,, \qquad \qquad 
    v^{(3)}_k=z_k\pm\frac{\ri\pi}{2}\,.
\end{aligned}    
\end{equation}
Note that the third `type of roots' $v_k^{(3)}$ form two-strings. By inserting the above form of roots into the logarithmic form of the Bethe equations, we obtain the following counting functions for their real parts $x_k$, $y_k$ and $z_k$
\begin{equation}
    \begin{aligned}
    c^x(x)=&-L \, \psi(x,\tfrac{\alpha-\gamma}{2})+L\,\psi\left(x,\tfrac{\alpha+\gamma}{2}\right)-\sum^{M_x}_{m=1}\, \psi\left(\tfrac{1}{2}(x-x_m),\tfrac{\gamma}{2}\right)\\&+\sum^{M_y}_{m=1}\, \phi\left(\tfrac{1}{2}(x-y_m),\tfrac{\gamma}{2}\right)-\sum^{M_z}_{m=1}\, \psi\left(x-z_m,\gamma\right)\,,\\
    c^y(x)=&-L \, \psi(x,\tfrac{\alpha-\gamma}{2})+L\,\psi\left(x,\tfrac{\alpha+\gamma}{2}\right)+\sum^{M_x}_{m=1}\, \phi\left(\tfrac{1}{2}(x-x_m),\tfrac{\gamma}{2}\right)\\&-\sum^{M_y}_{m=1}\, \psi\left(\tfrac{1}{2}(x-y_m),\tfrac{\gamma}{2}\right)-\sum^{M_z}_{m=1}\, \psi\left(x-z_m,\gamma\right)\,,\\
    c^z(x)=&-L \, \psi(x,\tfrac{\alpha-\gamma}{2})+L\,\psi\left(x,\tfrac{\alpha+\gamma}{2}\right)-\sum^{M_x}_{m=1}\, \psi\left(\tfrac{1}{2}(x-x_m),\tfrac{\gamma}{2}\right)\\&+\sum^{M_y}_{m=1}\, \phi\left(\tfrac{1}{2}(x-y_m),\tfrac{\gamma}{2}\right)-\sum^{M_z}_{m=1}\, \psi\left(x-z_m,\gamma\right)\,.
\end{aligned}
\end{equation}
In the above formula, the total number of roots is constrained to $M_x+M_y+2M_z=2L$. In terms of the counting functions, the Bethe equations become
\begin{align}\label{Log_BAE}
    c^x(x_k)=2\pi I^x_k\,,\quad   c^y(y_k)=2\pi I^y_k\,,\quad   c^z(z_k)=2\pi I^z_k\,.
\end{align}
where the $I^{x,y,z}_k$ are (half-)integers for $M^{x,y,z}$ (odd) even.

The corresponding root densities $\rho^a(x)=\partial c^a(x)/\partial x$ ($a=x,y,z$) resemble $\rho_1(x)$ and $\rho_2(x)$ obtained for the periodic model, see Eq.~(3.9) in \cite{FrSe14}: while the densities $\rho^{x,y}$ of the real parts of the type one (type 2) $v^{(1)}$ ($v^{(2)}$) turn out to be the same as $\rho_1$, the density $\rho^{z}$ of the real centres of the two-strings coincides with $\rho_2$.
Replacing the sum over the Bethe roots in \eqref{uhu} by an integral over the root densities, the energy density \eqref{eInf} is recovered.

\begin{figure}
    \centering
    \includegraphics[width=0.7\linewidth]{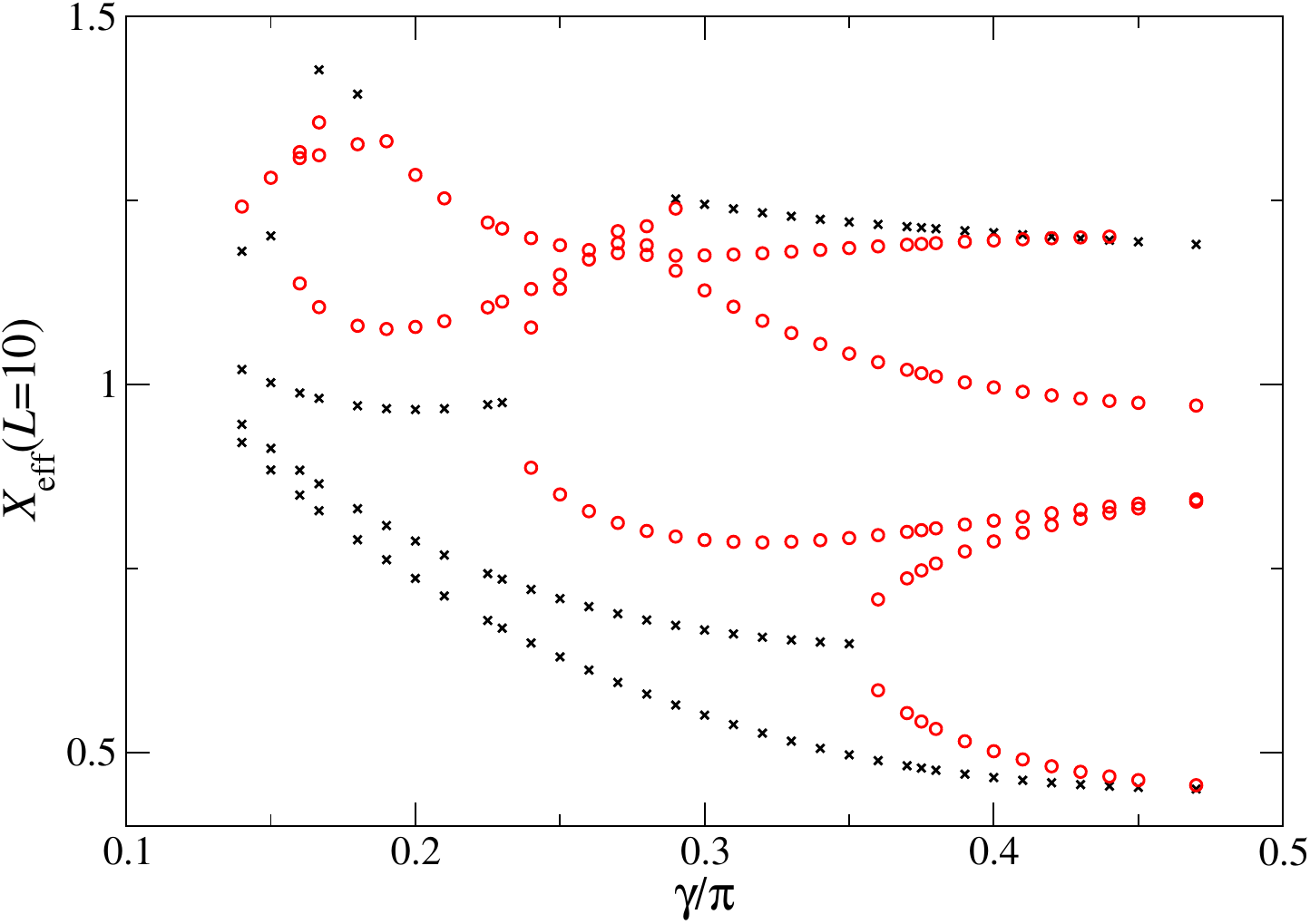}
    \caption{Lowest scaling dimensions (\ref{Xeff}) in phase III: red circles are derived from the lowest real eigenvalues of (\ref{End_H-Op}) for $\alpha=\frac{\pi}{2}$ and $L=10$, i.e.\ $20$ lattice sites. Black crosses from the real  part of complex eigenvalues. }
    \label{fig:spec10}
\end{figure}

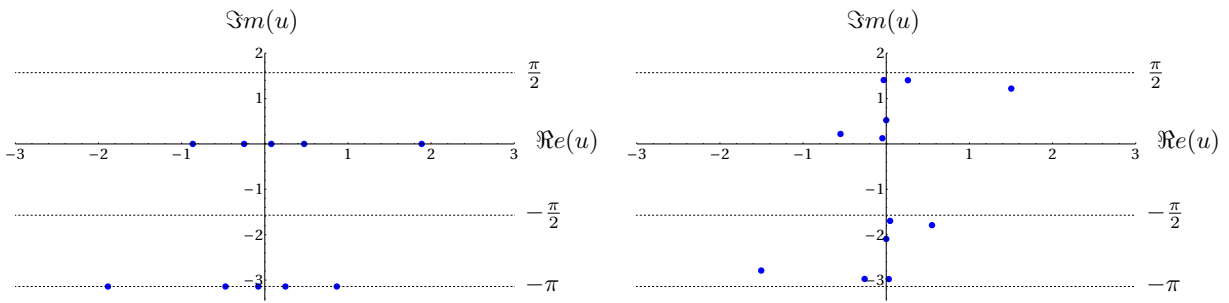
\begin{figure}
    \centering
    \begin{minipage}[b]{.49\linewidth}
    \scalebox{0.93}{
\begin{tikzpicture}
\node at (0,0) {\includegraphics[width=0.9\linewidth]{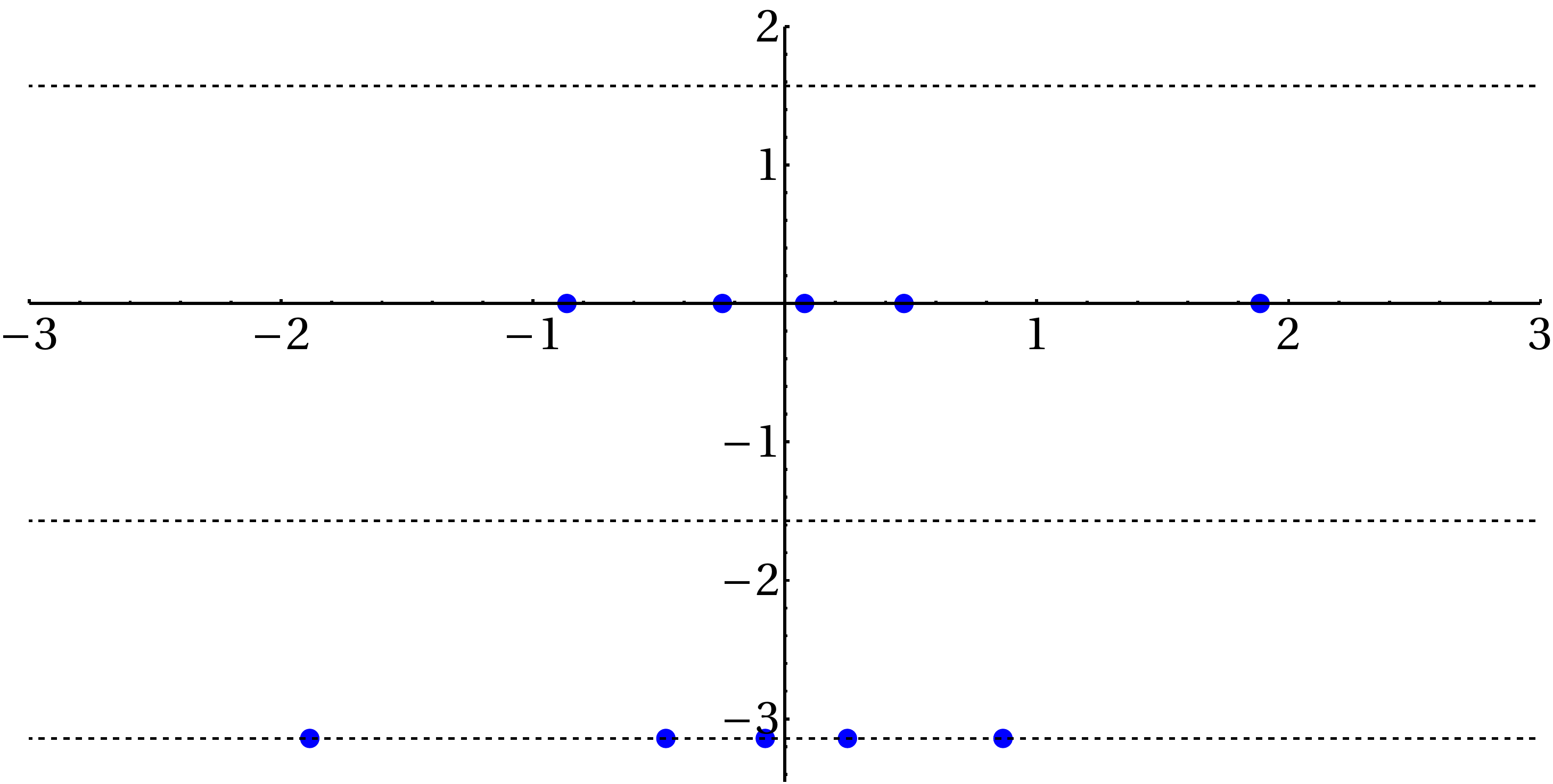}};
\node at (0.00,2.15)   {\small$\Im m(u)$};
\node[anchor=west] at  (3.75,0.45)  {\small$\Re e(u)$};

\node[anchor=west] at (3.6,1.4)   {\small$\tfrac{\pi}{2}$};
\node[anchor=west] at (3.6,-0.6)   {\small$-\tfrac{\pi}{2}$};
\node[anchor=west] at (3.6,-1.6)  {\small$-\pi$};
\end{tikzpicture}
}
 \end{minipage}
 \begin{minipage}[b]{.49\linewidth}
      \scalebox{0.93}{
\begin{tikzpicture}
\node at (0,0) {\includegraphics[width=0.9\linewidth]{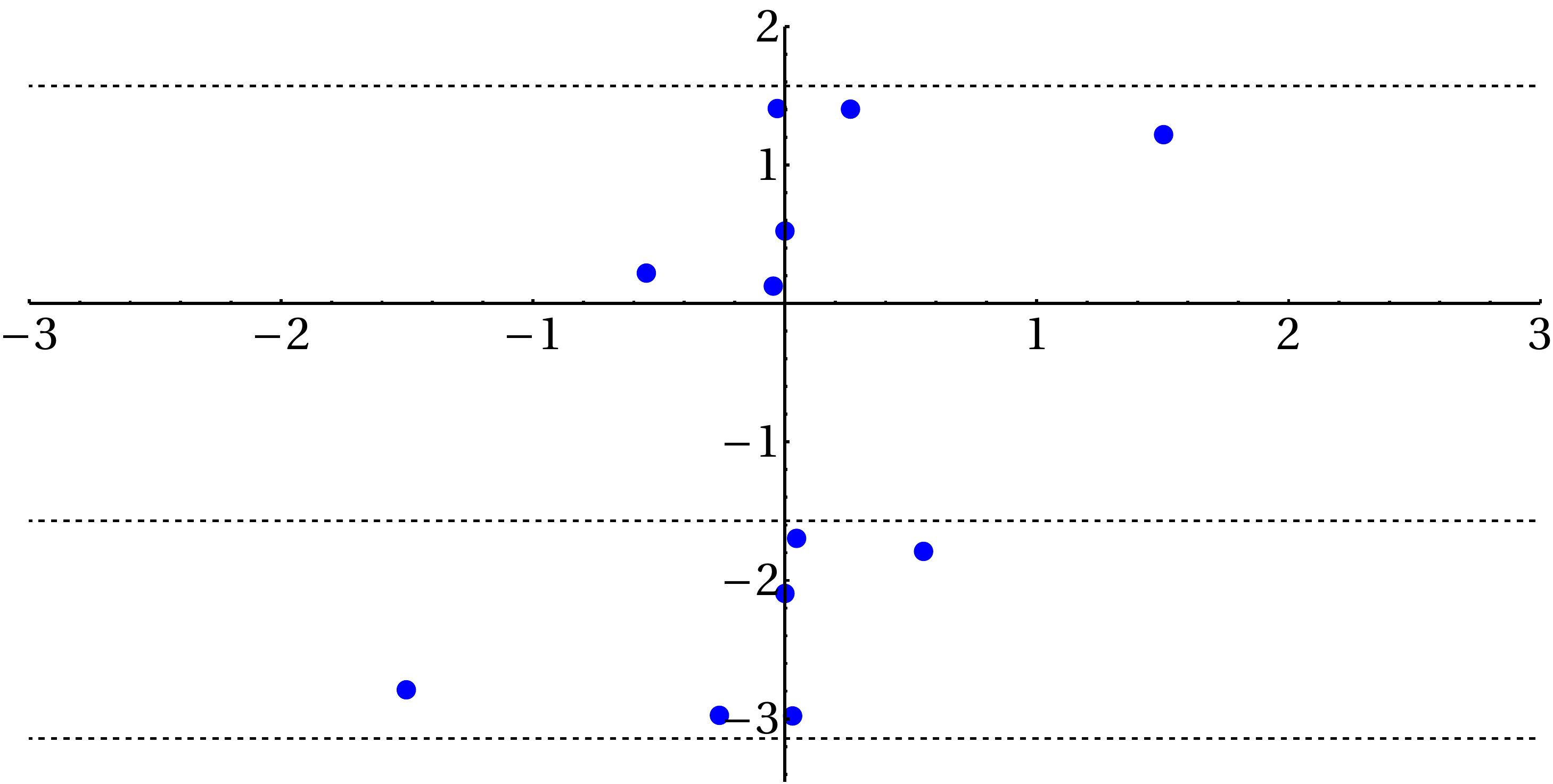}};
\node at (0.00,2.15)   {\small$\Im m(u)$};
\node[anchor=west] at  (3.75,0.45)  {\small$\Re e(u)$};

\node[anchor=west] at (3.6,1.4)   {\small$\tfrac{\pi}{2}$};
\node[anchor=west] at (3.6,-0.6)   {\small$-\tfrac{\pi}{2}$};
\node[anchor=west] at (3.6,-1.6)  {\small$-\pi$};
\end{tikzpicture}
}
\end{minipage}
    \caption{The left (right) plot displays the dual Bethe-root configuration of the ground state for the phase III in the complex $u$-plane for $2L=10$, ($2L=12$), $\gamma=\frac{\pi}{6}$ and $\alpha=\tfrac{\pi}{2}$.}
    \label{BR_GS_A_1_Crossing}
\end{figure}

In our analysis of small system sizes we observe a sequence of level crossings and the appearance of complex energies when the anisotropy $\gamma$ is  changed, see Fig.~\ref{fig:spec10}.
Moreover, the study of small systems does not yield reliable support for the construction of RG trajectories based on our Bethe hypothesis (\ref{Bethe_Hypothesis}) as the root patterns may change drastically when $L$ is varied, see Figure \ref{BR_GS_A_1_Crossing}.

Based on the diagonalization of the Hamiltonian for system sizes up to $2L\le28$ using the Arnoldi-Krylov method \cite{Kryl31,Arno51} we have identified the root configurations for all low energy states. It turns out that only states parameterized by root configurations (\ref{Bethe_Hypothesis}) with quantum numbers $I^x_k$, $I^y_k$ shifted against each other are realized on the lattice, e.g.\ 
\begin{equation}
\begin{aligned}
    I^x_k&\approx-\frac{M_x-1}{2}+\frac{1}{2},-\frac{M_x-1}{2}+\frac{3}{2},\dots, +\frac{M_x-1}{2}+\frac{1}{2}\\
    I^y_k&\approx-\frac{M_y-1}{2}-\frac{1}{2},-\frac{M_y-1}{2}+\frac{1}{2},\dots, +\frac{M_y-1}{2}-\frac{1}{2}\\
    I^z_k&\approx-\frac{M_z-1}{2}-\frac{1}{2},-\frac{M_z-1}{2}+\frac{1}{2},\dots, +\frac{M_z-1}{2}-\frac{1}{2}
\end{aligned}    
\end{equation}
for the Bethe root configuration displayed in the right panel of Figure \ref{Dual_Vs_Non_Dual}.

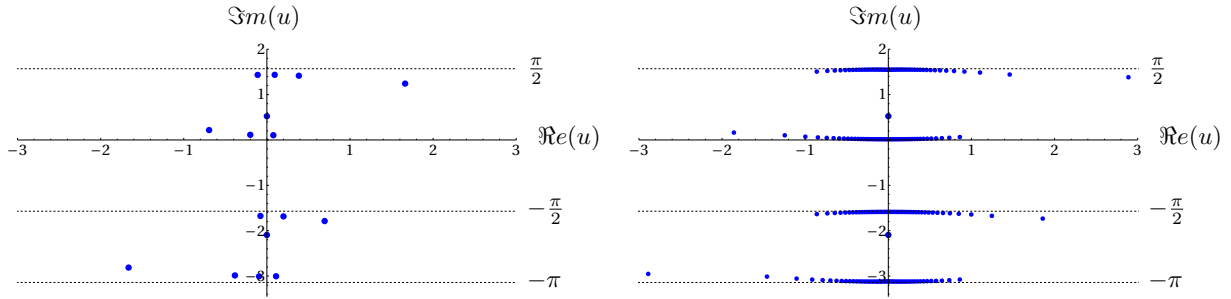
\begin{figure}[t]
    \centering
    \begin{minipage}[b]{.49\linewidth}
    \scalebox{0.93}{
\begin{tikzpicture}
\node at (0,0) {\includegraphics[width=0.9\linewidth]{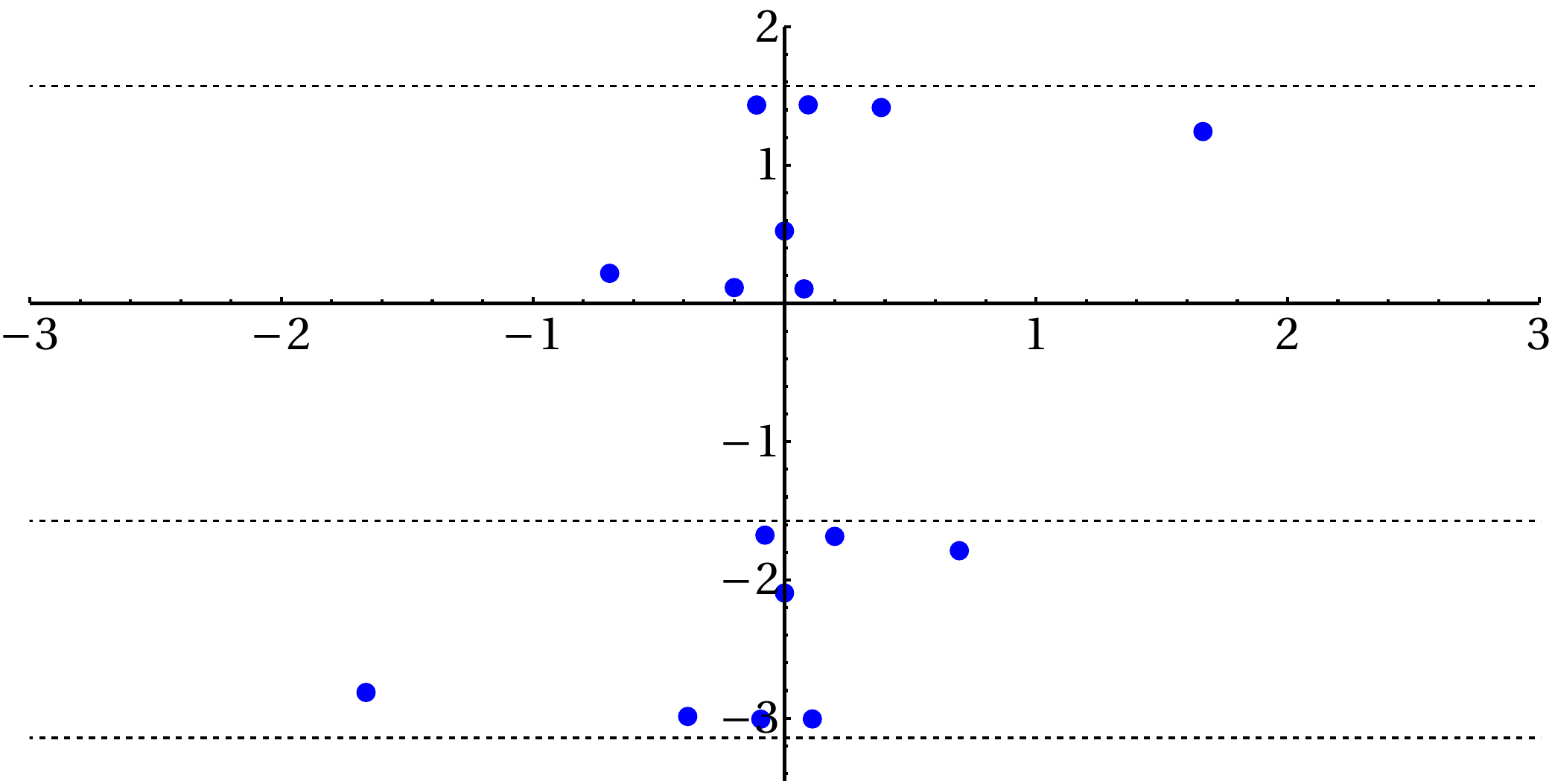}};
\node at (0.00,2.15)   {\small$\Im m(u)$};
\node[anchor=west] at  (3.75,0.45)  {\small$\Re e(u)$};
\node[anchor=west] at (3.6,1.4)   {\small$\tfrac{\pi}{2}$};
\node[anchor=west] at (3.6,-0.6)   {\small$-\tfrac{\pi}{2}$};
\node[anchor=west] at (3.6,-1.6)  {\small$-\pi$};
\end{tikzpicture}
}
 \end{minipage}
 \begin{minipage}[b]{.49\linewidth}
      \scalebox{0.93}{
\begin{tikzpicture}
\node at (0,0) {\includegraphics[width=0.9\linewidth]{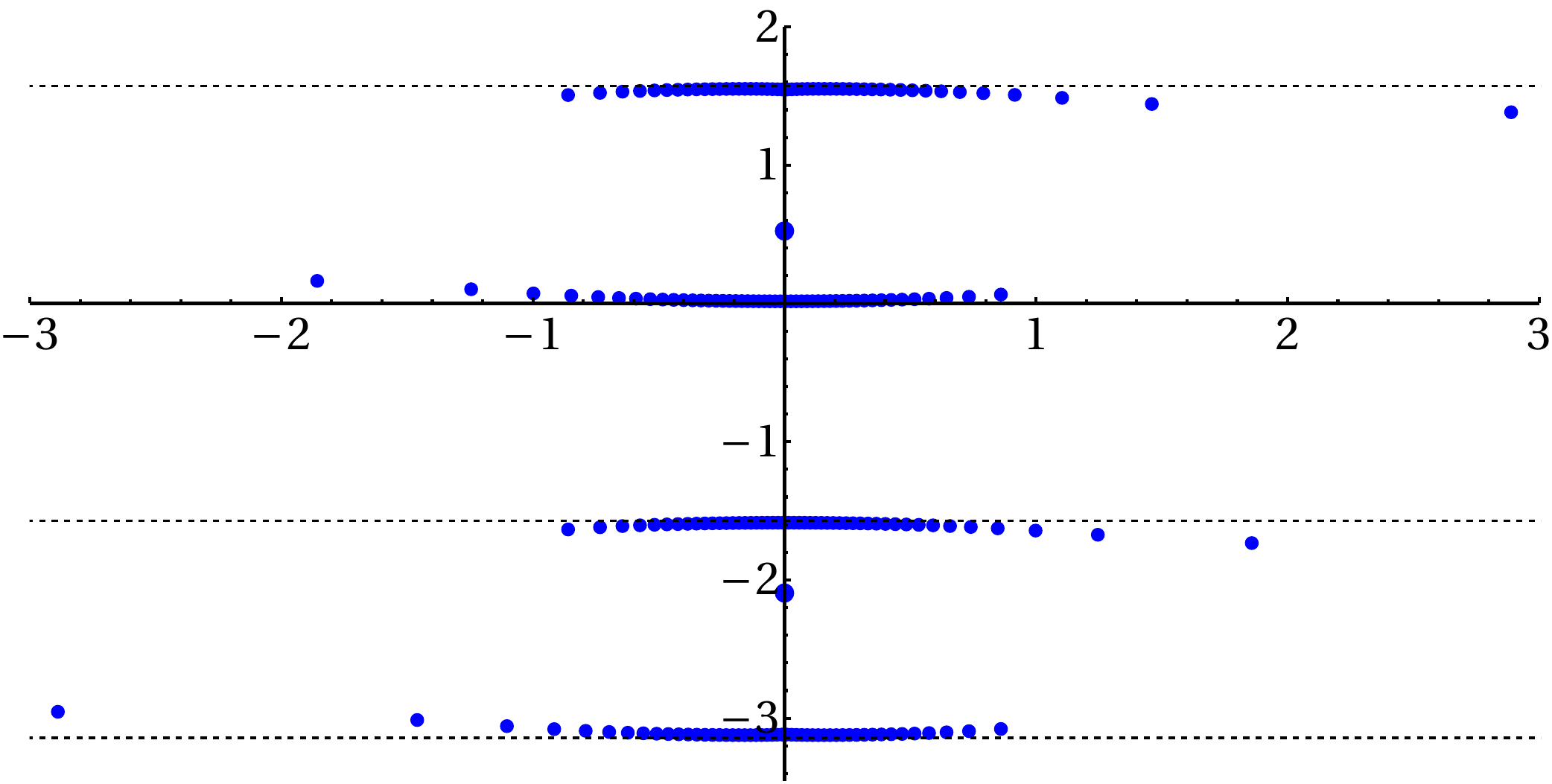}};
\node at (0.00,2.15)   {\small$\Im m(u)$};
\node[anchor=west] at  (3.75,0.45)  {\small$\Re e(u)$};
\node[anchor=west] at (3.6,1.4)   {\small$\tfrac{\pi}{2}$};
\node[anchor=west] at (3.6,-0.6)   {\small$-\tfrac{\pi}{2}$};
\node[anchor=west] at (3.6,-1.6)  {\small$-\pi$};
\end{tikzpicture}
}
\end{minipage}
    \caption{Left (right) plot displays the dual Bethe-root configuration of the ground state for $\gamma=\frac{\pi}{6}$,   $\alpha=\tfrac{\pi}{2}$ in phase III in the complex $u$-plane for $2L=16$ ($2L=200$).}
    \label{BR_GS_Ex_A_1}
\end{figure}

\begin{figure}
    \centering
    \begin{minipage}[b]{.49\linewidth}
    \scalebox{0.93}{
\begin{tikzpicture}
\node at (0,0) {\includegraphics[width=0.9\linewidth]{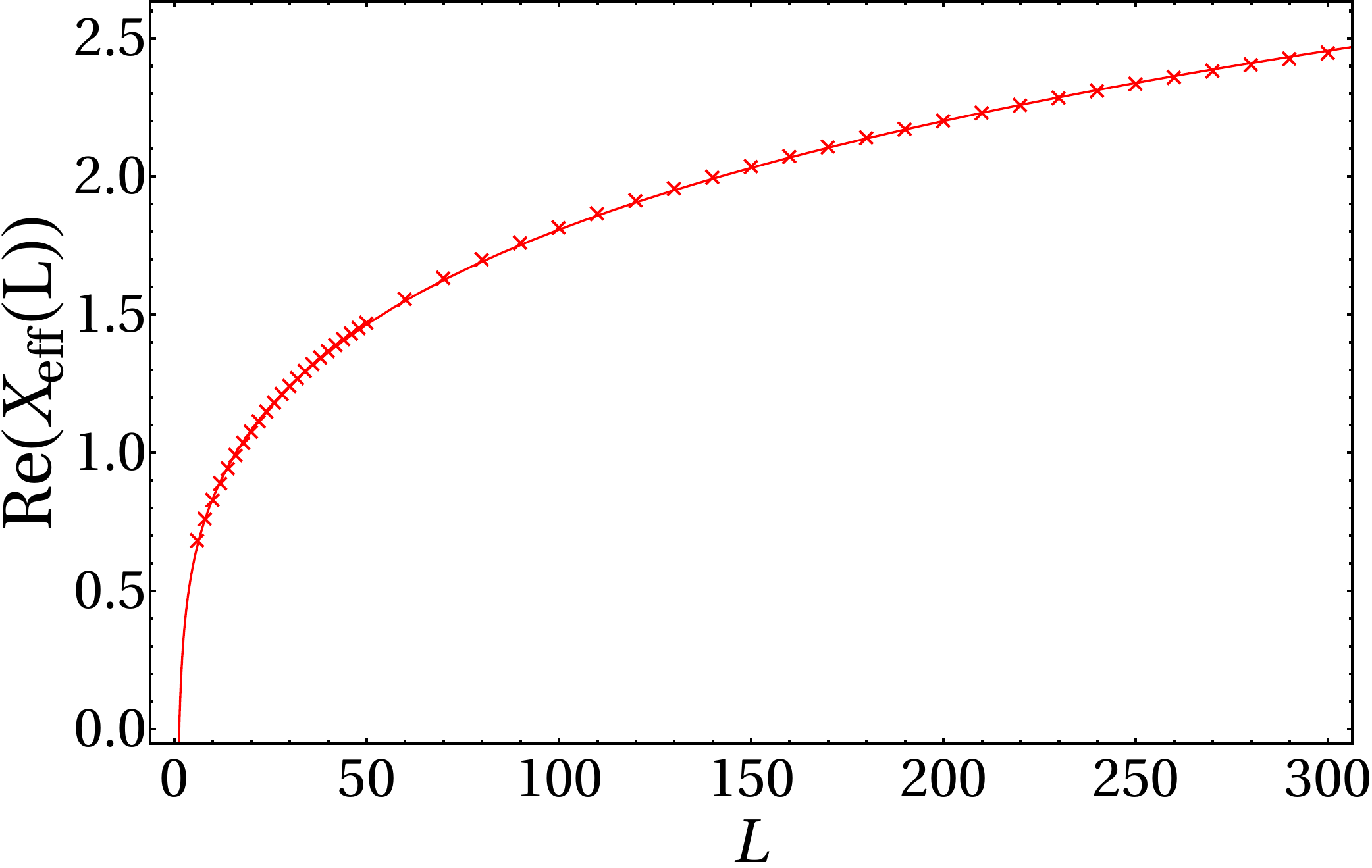}};
\end{tikzpicture}
}
 \end{minipage}
 \begin{minipage}[b]{.49\linewidth}
      \scalebox{0.93}{
\begin{tikzpicture}
\node at (0,0) {\includegraphics[width=0.9\linewidth]{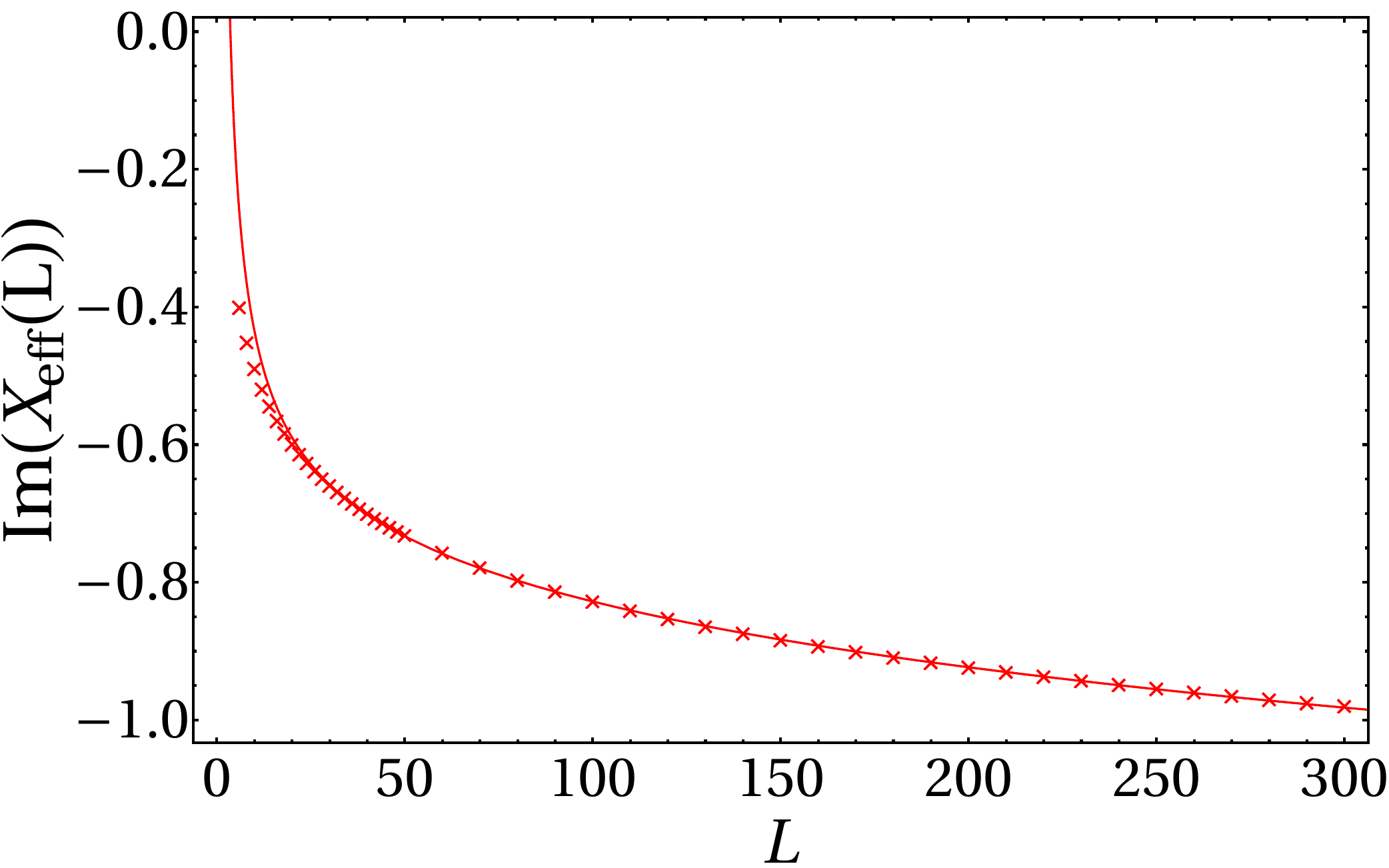}};
\end{tikzpicture}
}
\end{minipage}
    \caption{Left (right) plot displays the real (imaginary) part of the effective scaling dimensions of the ground state for $\gamma=\frac{\pi}{6}$ and $\alpha=\frac{\pi}{2}$ as function of the system size $L$. The corresponding root configurations are illustrated in Figure \ref{BR_GS_Ex_A_1}. The solid lines are fits to the function $a_1+a_2 \log^2(L)+a_3L^{-1}$.}
    \label{RG_GS_A_1}
\end{figure}

\begin{figure}[t]
    \centering
    \begin{minipage}[b]{.49\linewidth}
    \scalebox{0.99}{
\begin{tikzpicture}
\node at (0,0) {\includegraphics[width=0.9\linewidth]{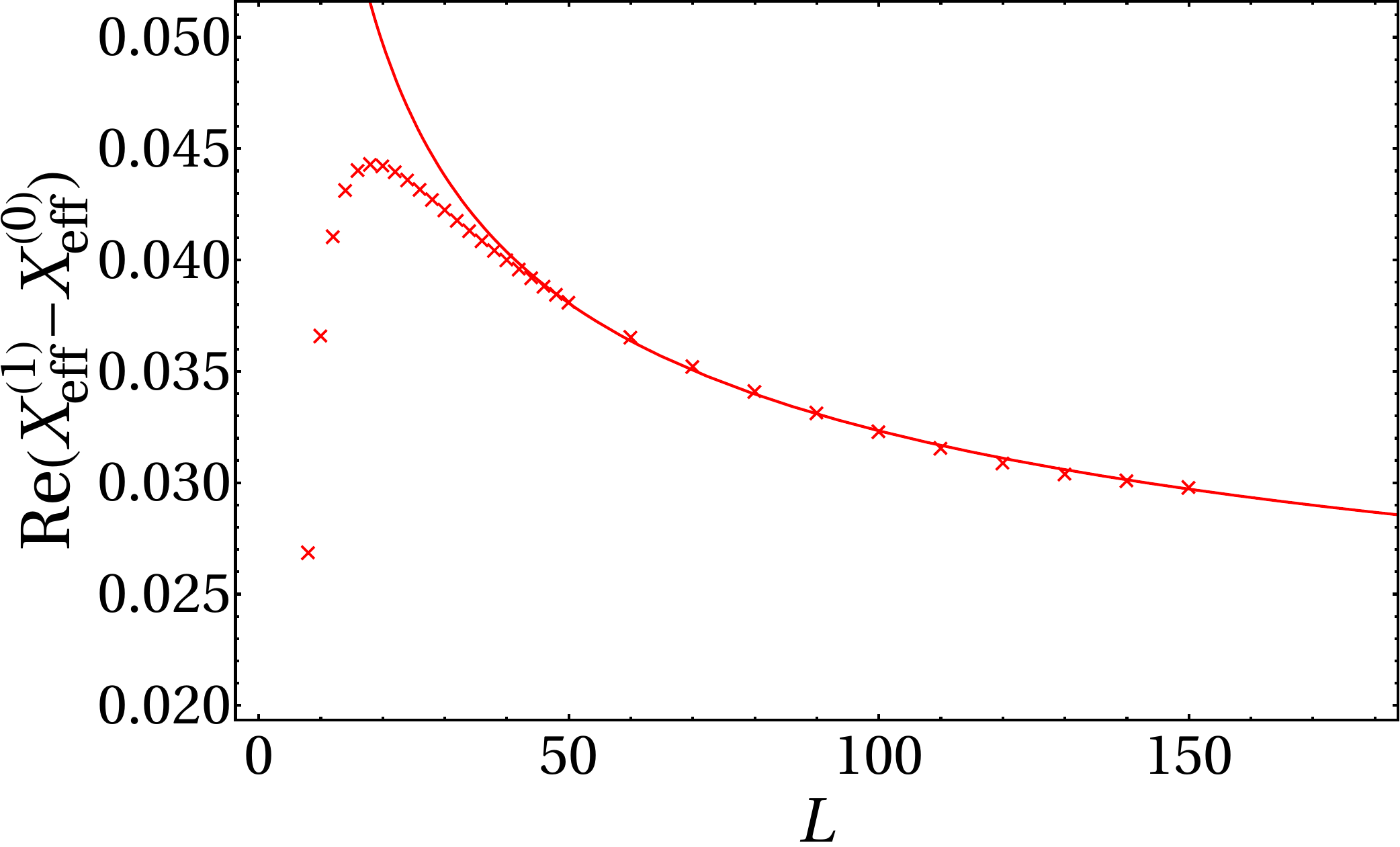}};
\end{tikzpicture}
}
 \end{minipage}
 \begin{minipage}[b]{.49\linewidth}
      \scalebox{0.99}{
\begin{tikzpicture}
\node at (0,0) {\includegraphics[width=0.9\linewidth]{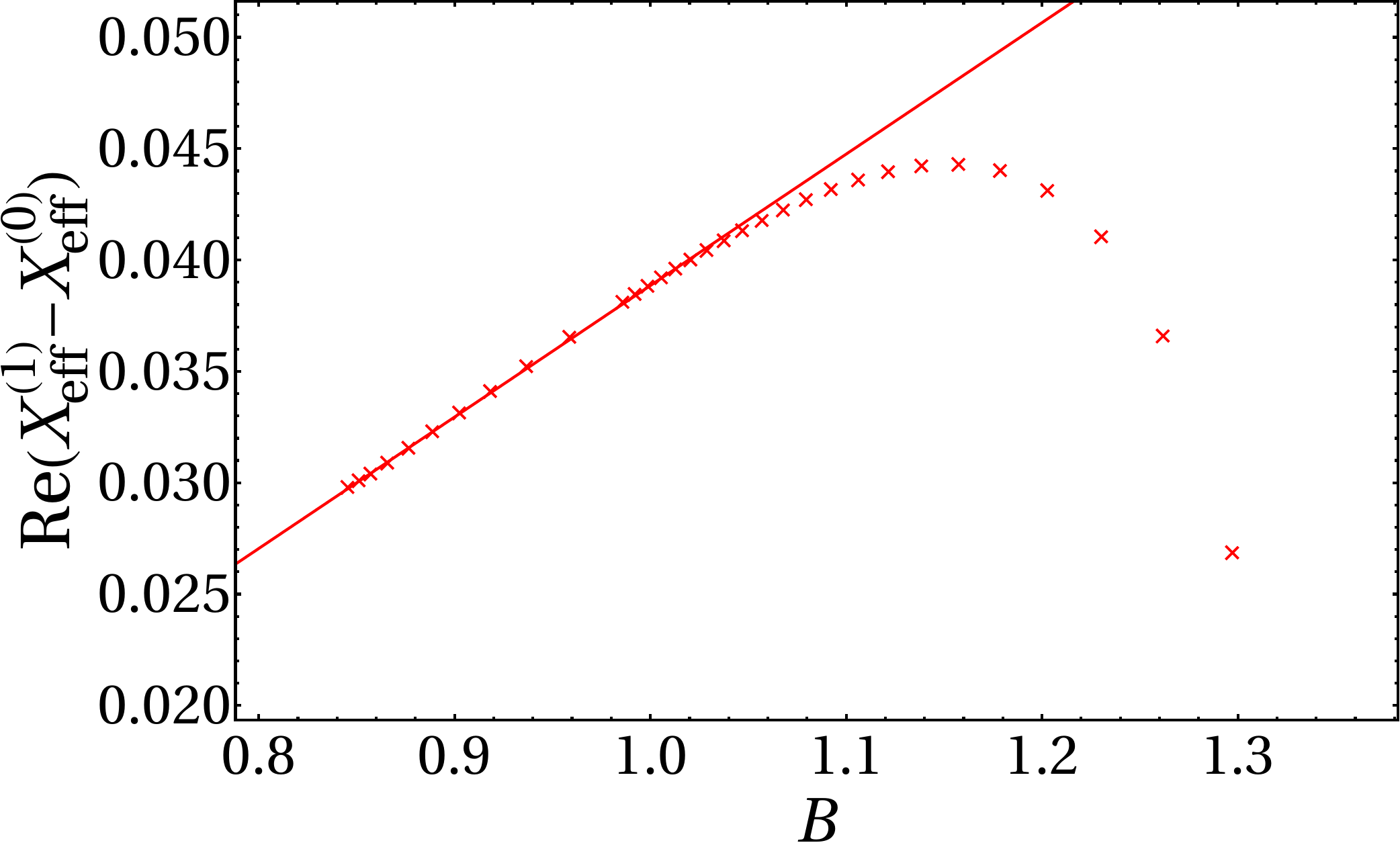}};
\end{tikzpicture}
}
\end{minipage}
    \caption{The real part of the difference between the effective scaling dimensions of the first excited state and the ground state is shown for $\gamma=\tfrac{\pi}{6}$ and $\alpha=\frac{\pi}{2}$. Left: For large system sizes the gaps close logarithmically, the solid line is a fit to $a_1/\log(L)$. Right: For large system sizes (small $B$) the gaps become a linear function of the quasi-momentum \eqref{oxnsh}. }
    \label{GAPS}
\end{figure}

We have investigated the scaling behaviour of such states on the lattice, e.g.\ the ground state for $L=8$ and its extension to $L=100$ displayed in Figure \ref{BR_GS_Ex_A_1}. 
They show corrections to scaling which increase as $(\log L)^2$ with the system size, see Figure~\ref{RG_GS_A_1}.  This holds also for the imaginary parts of complex energies, such that the appearance of complex energies cannot be argued to be a finite-size effect. 
This behaviour is a signature of the profound influence of antidiagonal BC on the critical behaviour in the (black hole) phase III.  Since a finite study based on the CFT prediction (\ref{Xeff}) is not possible under such circumstances we have considered the finite size scaling of the energy differences between the ground state and the first excited state, see Figure \ref{GAPS}.
The energy gap appears to close as $\sim {1}/{L\log(L)}$. Further, the closing of the energy gap seems to be linearly related to the quasi-momentum operator (\ref{QI}). This might signal the remains of the continuous component of the finite-size spectrum observed in phase III for quasi-periodic BC, although the scaling of the ground state is very different.

In summary, taking the RG trajectories seeded by the ground state and first excitation for $L=8$ as a reference the corrections to scaling diverge while the energy gap to the first excitation closes logarithmically: 
\begin{equation}
\label{scalingIII}
\begin{aligned}
    L(E_0 - L\epsilon_\infty)&\sim a_1+a_2\log^2(L) \,,\\
    L(E_{1}-E_{0})&\sim \Tilde{a}_1+\frac{\Tilde{a}_2}{\log(L)}\,,
\end{aligned}    
\end{equation}
where $a_i,\,\Tilde{a}_i\in\mathbb{C}$.
Speculating about the origin of the logarithmic growth of the subleading corrections to $E_0$ one may consider two possible scenarios:
(a) the corresponding RG trajectory does not connect to the true ground state (or even belong to the low energy sector) with bulk energy (\ref{eInf}) in the scaling limit.
(b) The RG trajectory does describe the ground state of the model but the antidiagonal boundary conditions  lead to a marginal relevant perturbation of the CFT describing the scaling limit for quasi-periodic boundary conditions (with diagonal twist).
As for (a) we have not observed any evidence for the existence of another set of RG trajectories starting from the small $L$ low energy spectrum with decreasing corrections to scaling as $L\to\infty$. Hence, to rule out the existence of such a set the finite size analysis of the spectrum has to be extended significantly.  As discussed above, such an analysis has to address the problem of a large number of level crossings (with the system size) interfering with the construction of the corresponding RG trajectories.
In the case (b) the finite size scaling (\ref{scalingIII}) describes the RG flow in the vicinity of the (unstable) fixed point under the perturbation by the antiperiodic boundary conditions.  Since the integrable model does not contain any further parameter, however, it will not be possible to identify the corresponding stable fixed point within this setting.

In both cases the characterization of the scaling limit in phase III requires substantial additional effort.  Therefore, we leave this problem for a future research project on its own.  Our numerical data for the Bethe root configurations of the ground state and first excitation used in Figures \ref{BR_GS_Ex_A_1}, \ref{RG_GS_A_1} and \ref{GAPS} are available online \cite{DATA}.


\begin{acknowledgments}
    The authors thank Gleb Kotousov for valuable discussions. This work is supported by the Deutsche Forschungsgemeinschaft (DFG) under grant No. Fr 737/9-2. Part of the numerical work has been performed on the LUH computer cluster, which is funded by the Leibniz Universität Hannover, the Lower Saxony Ministry of Science and Culture and the DFG.
\end{acknowledgments}

\appendix

\section{Matrix elements of the Q-operator \label{QQ}}
In this appendix, we present our result for the Baxter Q-operator. We have found that the matrix elements of the Q-operator in the tensor product basis $\ket{j} = \ket{\mathbf{s}(j)}=\otimes_{k=1}^{2L}\ket{s_{k}(j)}$, with $s_k(j)=\pm1$ and $\sigma_k^z\ket{j}=s_k(j)\ket{j}$, take the following form
\begin{align}
    \left[\mathbbm{Q}(u)\right]_{i,j}=\frac{Q^E_{i,j}(u)}{Q^E_{j,j}(u)}\,.\label{afnmeifQ}
\end{align}
Here $Q^E_{i,j}(u)$ has been defined to be 
\begin{align}
\label{afnmeif}
    Q^E_{i,j}(u)= \prod_{k=1}^{2L} h_{k,s_k(i)}(\mathbf{s}(j))\,\re^{u s_k(j)/2}\,
\end{align}
where
\begin{equation*}
\begin{aligned}
    h_{k,s}(\mathbf{t}) &= \begin{cases}
        1 & \text{for~} s=-1\\
        r_k(\mathbf{t})\, \re^{-ut_k}\,\re^{\delta_k\, t_k}  & \text{for~} s=+1
    \end{cases}\,,\\
    r_k(\mathbf{t}) &= \re^{-(\ri\gamma/2)\sum_{j=1}^k t_j}\, \re^{(\ri\gamma/2)\sum_{j=k+1}^{2L} t_j}\,.
\end{aligned}
\end{equation*}

Note that this implies a normalisation of the Q-operator different from (\ref{ddskdmk1}), i.e.\ $\mathbbm{Q}(u\to-\infty)=\mathbbm{1}$ with eigenvalues
\begin{align}
    Q(u)=\prod_{j=1}^{2L}\left(1-\re^{u-v_j+\ri\gamma/2} \right)\,,
\end{align}
where $v_j$ are the Bethe roots solving \eqref{ddskdmk2}.
In this normalisation, the TQ-relation \eqref{dqeikjfi} gets modified by two additional phase factors 
\begin{align}
\label{eq:TQ_renorm}
    \mathbbm{t}(u)\mathbbm{Q}(u)=\re^{\ri L \gamma }\prod^{2L}_{j=1}\sinh(u-\delta_j+\ri\gamma) \mathbbm{Q}(u-\ri\gamma)-\re^{-\ri L \gamma }\prod^{2L}_{j=1}\sinh(u- \delta_j)\mathbbm{Q}(u+\ri\gamma)\,.
\end{align}

Let us briefly comment on the relationship to the Q-operator introduced in Ref.~\cite{BBOY95}. While \eqref{afnmeif} is a straightforward generalisation of the corresponding object in that work to the inhomogeneous case, the expression \eqref{afnmeifQ} is new. In \cite{BBOY95} the Q-operator satisfying \eqref{dqeikjfi} has been defined as $\widetilde{\mathbbm{Q}}(u)=Q^E(u)[Q^E(u_0)]^{-1}$, where $Q^E$ is the matrix built out of $Q^E_{i,j}$ (\ref{afnmeif}) and $u_0$ is an arbitrary fixed value of the spectral parameter.  
Surprisingly, we have found that normalising each matrix element $Q^E_{i,j}(u)$ with the associated  diagonal elements $Q^E_{j,j}(u)$ as in \eqref{afnmeifQ} leads to a Q-operator which commutes with itself and the transfer matrix for different values of the spectral parameter and satifies the TQ-relation (\ref{eq:TQ_renorm}).  We have checked this fact numerically for small system sizes.  Since \eqref{afnmeifQ} does not involve any additional matrix operations on $Q^E(u)$ this expression is particularly useful for the numerical computation of the Bethe root configurations.

\end{document}